%% file: main.tex
\documentclass[acmtog,nonacm]{acmart}

\acmJournal{TOG}
\usepackage{amsmath}
\usepackage{graphics}
\usepackage{color}
\usepackage{graphicx}
\usepackage{placeins} 
\usepackage{makecell} 
\usepackage{amssymb}
\usepackage{multirow}
\usepackage{enumerate}
\usepackage{bbding}
\usepackage{pifont} 
\usepackage{wasysym} 
\usepackage{utfsym} 
\usepackage{fontawesome} 

\usepackage[percent]{overpic} 

\newcommand{\cmark}{\checkmark}
\newcommand{\xmark}{×}
\newcommand{\chmark}{{\cmark}\textsuperscript{{\kern-0.6em\tiny\xmark}}}

\newcommand{\SQ}[1]{{#1}}
\newcommand{\ZY}[1]{{\color{cyan}ZY: #1}}
\newcommand{\XR}[1]{{#1}}

\setcopyright{acmcopyright}\acmJournal{TOG}
\acmYear{2022}\acmVolume{41}\acmNumber{6}\acmArticle{228}\acmMonth{12} \acmDOI{10.1145/3550454.3555443}

\begin{document}

\title{RFEPS: Reconstructing Feature-line Equipped Polygonal Surface
}

\author{Rui Xu}
\affiliation{  \institution{Shandong University} 
\country{China}}\email{xrvitd@163.com}

\author{Zixiong Wang}
\affiliation{  \institution{Shandong University}
\country{China}}\email{zixiong_wang@outlook.com}

\author{Zhiyang Dou}
\affiliation{  \institution{The University of Hong Kong}
\country{China}}\email{zhiyang0@connect.hku.hk}

\author{Chen Zong}
\affiliation{  \institution{Shandong University}
\country{China}}\email{zongchen@mail.sdu.edu.cn}

\author{Shiqing Xin}
\authornote{Co-corresponding authors: Shiqing Xin and Changhe Tu. }
\affiliation{  \institution{Shandong University}
\country{China}}\email{xinshiqing@sdu.edu.cn}

\author{Mingyan Jiang}
\affiliation{  \institution{Shandong University}
\country{China}}\email{jiangmingyan@sdu.edu.cn}

\author{Tao Ju}
\affiliation{  \institution{Washington University in St. Louis}
\country{USA}}\email{taoju@wustl.edu}

\author{Changhe Tu} 
\authornotemark[1]
\affiliation{  \institution{Shandong University}
\country{China}}
\email{chtu@sdu.edu.cn}







\input{abs}



\begin{teaserfigure}
  \centering
  \vspace{-4mm}
  \includegraphics[width=\textwidth]{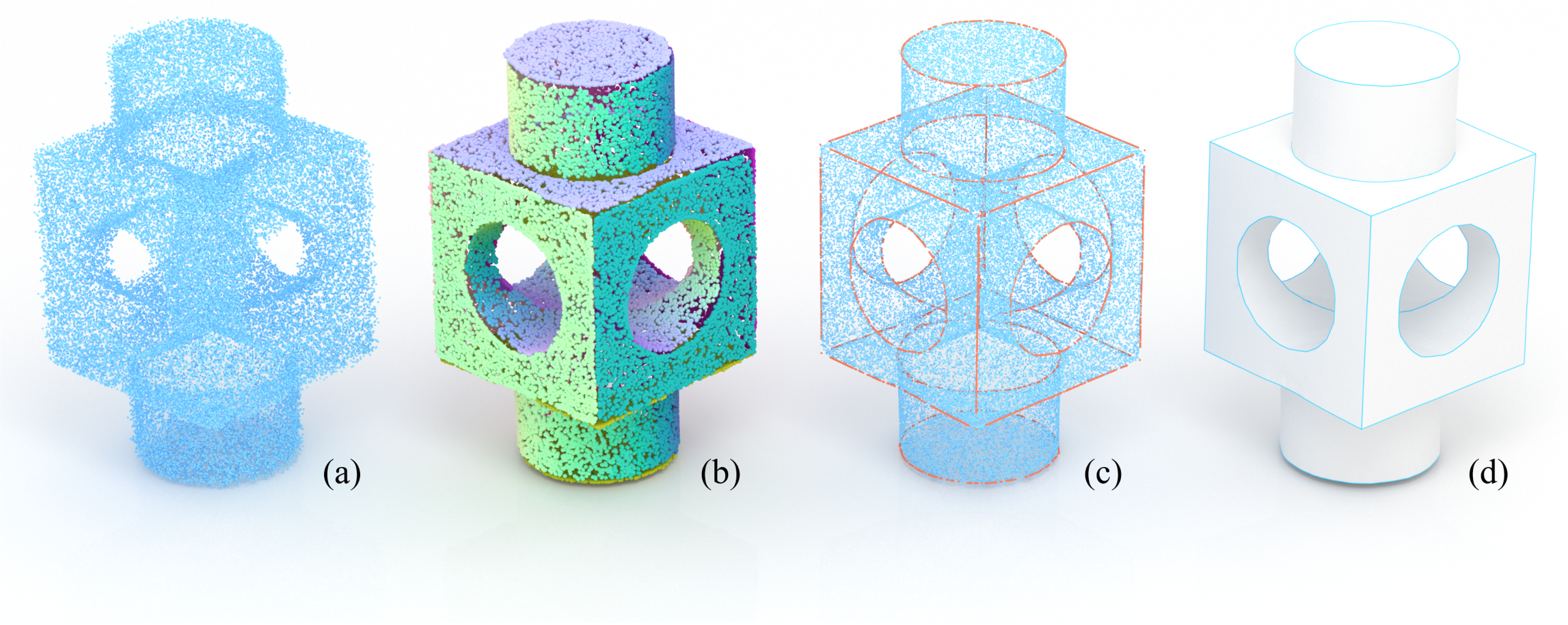}
  \vspace{-21mm}
  \caption{Reconstructing a polygonal surface with clean line-type features from a noisy point cloud.
  From left to right: (a)~The input point cloud is noisy and does not have reliable normal information. (b)~The point locations and normal vectors are optimized simultaneously such that 
  the resulting point cloud is as locally planar as possible (the normal vectors of the denoised point cloud are visualized in a color-coded style). 
  (c)~We augment the point set by predicting more points that are deemed to be located on potential geometry edges; See the points colored in red. 
  (d)~Based on a power-diagram decomposition restricted on the base surface (obtained by Poisson reconstruction),
  the resulting polygonal surface interpolates the augmented point set and naturally aligns with feature lines.}
  \label{fig:teaser}
\end{teaserfigure}

\maketitle

\input{introduction}

\input{RelatedWork}
\input{Method}

\input{Results}

\input{conclusion}





\FloatBarrier
\bibliographystyle{ACM-Reference-Format}
\bibliography{sample-base}


\end{document}

%% file: abs.tex
\begin{abstract}
Feature lines are important geometric cues in characterizing the structure of a CAD model. Despite great progress in both explicit reconstruction and implicit reconstruction, it remains a challenging task to reconstruct a polygonal surface equipped with feature lines, especially when the input point cloud is noisy and lacks faithful normal vectors. In this paper, we  develop a multistage algorithm, named {\em RFEPS}, to address this challenge. The key steps include (1)~denoising the point cloud based on the assumption of local planarity, (2)~identifying the feature-line zone by optimization of discrete optimal transport, (3)~augmenting the point set so that sufficiently many additional points are generated on potential geometry edges, and (4) generating a polygonal surface that interpolates the augmented point set based on restricted power diagram. We demonstrate through extensive experiments that RFEPS, benefiting from the edge-point augmentation and the feature preserving explicit reconstruction, outperforms state of the art methods in terms of the reconstruction quality, especially in terms of the ability to reconstruct missing feature lines. 
\end{abstract}

%% file: introduction.tex
\section{Introduction}
\begin{figure*}[!t]
	\centering
\vspace{-4mm}
 \includegraphics[width=\textwidth]{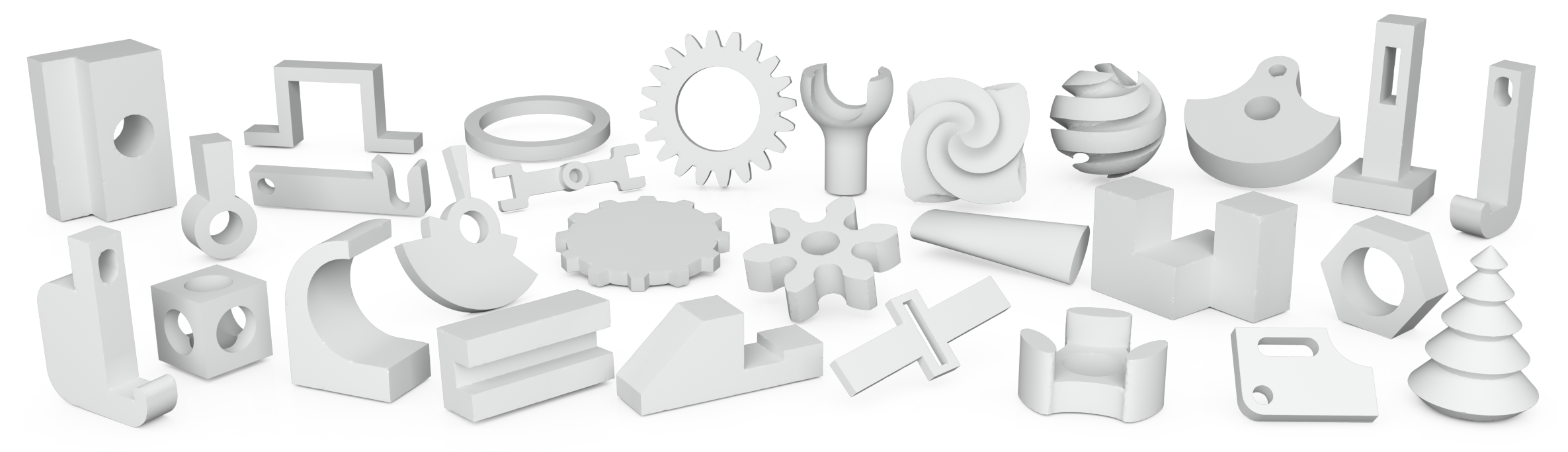}
\vspace{-8mm}
\caption{A gallery of reconstruction results by our RFEPS.}
	\label{FIG:results}
 	\vspace{-3mm}
\end{figure*}
Inferring geometry from a low-quality point cloud belongs to the reverse engineering category, which is a fundamental problem in computer-aided design~\cite{beniere2013comprehensive, willis2021engineering, birdal2019generic, li2019supervised, sommer2020primitect}. Line-type features are critical geometric cues and can help characterize the structure of a CAD model, and thus serve as a base for a wide range of semantic manipulation tasks~\cite{zhang2020blending,wu2005structure,kos2000methods}. In this paper, we 
take a noisy point cloud (the ground truth is a CAD model) without reliable normal vectors as the input and study how to recover a clean polygonal model exhibiting neat feature lines. 

The challenges are two-fold. 
On the one hand, the general input point cloud contains few points that are precisely located on the underlying geometry edges.
There are some point consolidation approaches~\cite{yu2018ec, EAR, CHENG2019101790} that aim at increasing the point density in the edge zone, but the added points do not define feature lines of the target polygonal surface due to imprecise positions or insufficient density. 
There are also some approaches~\cite{ma1998nurbs,dan2006algorithm,naik1988spline} that separate the given points into a collection of patches, followed by fitting a smooth surface to each patch. However, it is hard to find the best decomposition that reveals the structure of the underlying CAD model.
On the other hand, it is non-trivial to fully respect the line-type features during the mesh generation step.  
\SQ{The existing explicit reconstruction approaches lack effective techniques to prioritize the connections between edge points.}

In this paper, we introduce two separate techniques to deal with the above-mentioned difficulties. 
First, we observe that in the neighborhood of an edge point, the distribution of normal vectors can be represented as a combination of two independent sub-distributions with equal measure, and each sub-distribution is as simple as a Dirac delta function.
Based on this observation, we formulate the identification of the edge zone as a discrete optimal mass transport problem, which can help regularize point locations and normal vectors, and predict the edge points as well.
Second, we suggest using the restricted power diagram (RPD) to build connections between points. Compared with the restricted Voronoi diagram (RVD), the RPD not only has the nice feature of the RVD, but also allows  to set a larger weight for edge points,
thus encouraging edge points to naturally form feature lines with a higher priority than other points.

We propose a multistage algorithm named {\em RFEPS} to reconstruct a clean polygonal surface while manifesting the line-type features of CAD models.
Our algorithm begins with a denoising step, which the point locations and the normal vectors are optimized simultaneously based on the assumption that
a CAD-type point cloud tends to be locally planar;
See Figure~\ref{fig:teaser}(b). 
Next, we generate sufficiently many additional points that are on potential geometry edges; See  Figure~\ref{fig:teaser}(c).
Finally, we run the Poisson reconstruction solver to get a base surface and then compute the 
restricted power diagram~(RPD) to the base surface.
The dual of the RPD reports a feature preserving polygonal mesh that
interpolates the denoised and augmented point set;
See Figure~\ref{fig:teaser}(d).
Figure~\ref{FIG:results} shows a gallery of reconstruction results produced by our method.
More tests on both synthetic and raw-scan data are provided in
Section~\ref{sec:Results}.
We further demonstrate the utility of our approach in shape edit tasks; See Figure~\ref{FIG:CAD} in Section~\ref{sec:potential_application}.

To summarize, our contributions are threefold:
\vspace{-0.5mm} 
\begin{enumerate}
    \item We propose a multi-stage algorithm that 
    transforms a noisy point cloud of a CAD model into a watertight manifold polygonal surface with neat feature lines. 
    \item 
    We classify the points into nearby-edge points and off-edge points using a discrete optimal mass transport formulation,
    which enables us to generate sufficiently many additional points that 
    are accurately located on potential geometry edges. 
    \item We set a larger weight for the newly added edge points and then use the restricted power diagram to reconstruct a polygonal surface that interpolates the augmented point set and aligns  with the edges of the geometry.
\end{enumerate}

%% file: RelatedWork.tex
\section{Related Work}

\begin{figure*}[!htp]
	\centering
\vspace{-3mm}
\begin{overpic}
[width=.99\textwidth]{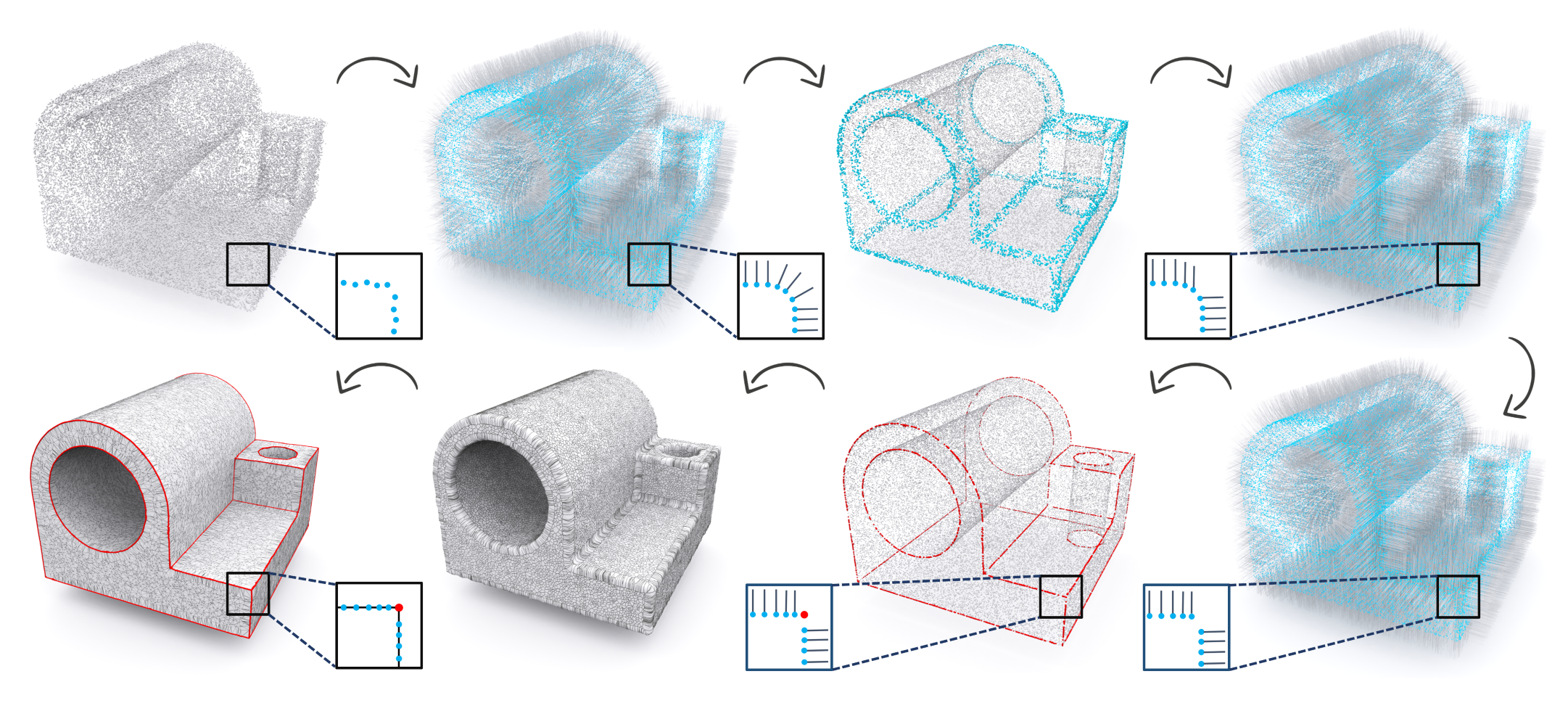}
\put(10,22.6){(a)}
\put(36,22.6){(b)}
\put(61,22.6){(c)}
\put(86,22.6){(d)}
\put(10,2){(h)}
\put(36,2){(g)}
\put(61,2){(f)}
\put(86,2){(e)}
\end{overpic}
\vspace{-5mm}
\caption{The pipeline of the proposed method. (a)~The noisy input point cloud. (b)~The denoised result by optimizing point locations and normal vectors simultaneously.
(c)~The edge zone detected by a discrete optimal transport formulation.
(d)~The normal vectors of the points in the edge zone are regularized.
(e)~The point locations are fine-tuned to adapt to the regularized normal vectors. 
(f)~Points in the edge zone are projected, point by point, onto the potential edge
such that there are sufficiently many additional edge points (colored in red). 
(g)~The restricted power diagram (RPD) on the base surface produced by the Poisson reconstruction solver. (h)~The dual of the RPD reports the final reconstructed polygonal surface 
that interpolates the augmented point set while naturally aligning feature lines along geometry edges. }
\vspace{-3mm}
\label{FIG:pipeline}
\end{figure*}
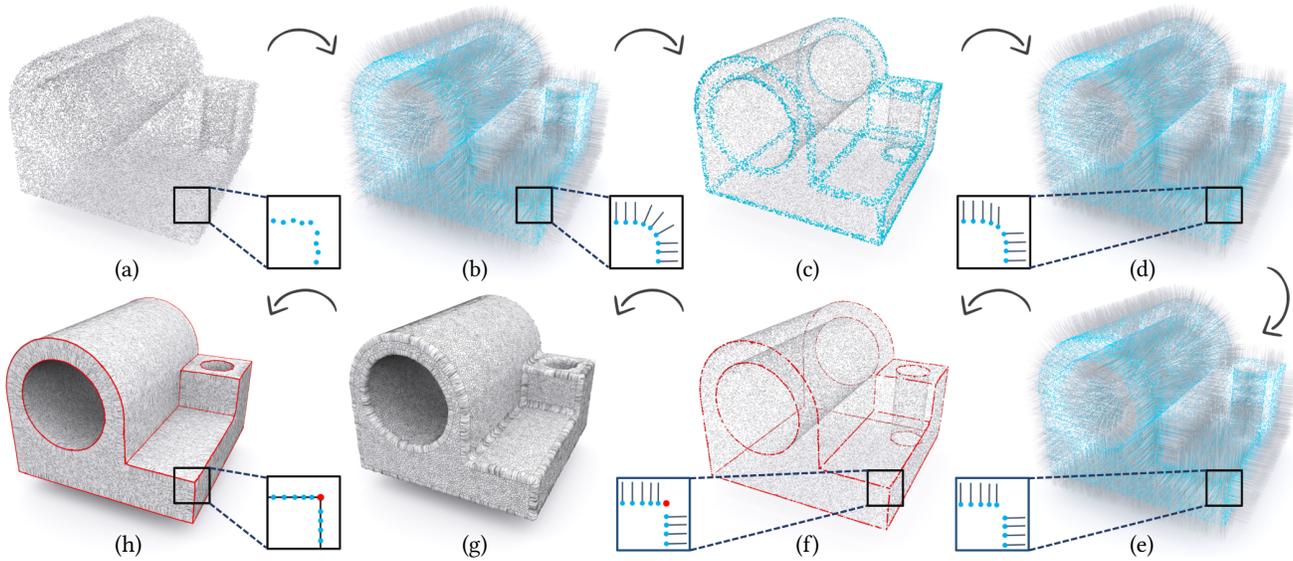

\paragraph{Point Cloud Consolidation}
There has been a large body of literature on consolidating point clouds in the past decade. 
Most of them assume that the underlying surface is globally smooth. 
Alexa et al.~\shortcite{alexa2001point}
suggested using moving least square (MLS)
to increase or decrease the density of the points,  allowing an adjustment of the
spacing among the points.
Lipman et al.~\shortcite{LOP} presented a locally optimal projection operator (LOP) 
that provides a second-order approximation to the underlying smooth surface, thus facilitating resampling of the original point data. 
Huang et al.~\shortcite{huang2009consolidation} 
proposed a weighted LOP for denoising and outlier removal from imperfect point data, producing a set of evenly distributed particles that accurately adheres to the captured shape.
However, the aforementioned approaches
are weak in dealing with sharp features, 
which motivates researchers to consider line-type features on point-sampled geometry during point consolidation~\cite{pauly2003multi, guennebaud2004real, rmls}.
Avron et al.~\shortcite{l1_sparse} 
defined an $l_1$-sparse optimization to denoise the input point cloud assuming that the target surface consists mainly of smooth patches,
where the residual of the objective function is strongly concentrated near sharp features.
Liao et al.~\shortcite{FLOP} 
took both spatial and geometric feature information into
consideration for feature-preserving approximation.
Huang et al.~\shortcite{EAR} presented Edge-Aware Resampling (EAR) that re-samples the points away from the edges and then progressively fills the gap between the planes. 
Although the point density near the edges is increased,
the added points are not precisely located on the edges.
Lu et al.~\shortcite{lu2020low} proposed a low-rank matrix approximation algorithm that can robustly estimate normals for both point clouds and meshes, which is helpful for edge-preserving upsampling.
Recently, Chen et al.~\shortcite{chen2021multiscale} presented a two-phase algorithm for extracting line-type features from point clouds.
\SQ{But the accuracy of the predicted edge points is decreased if noise exists.}

With the emergence of learning techniques, many data-driven approaches have been proposed to improve the robustness of edge  detection and the preservation of sharp edges.
To the best of our knowledge, EC-Net~\cite{yu2018ec} is the first learnable edge-aware method that
formulates a regression component to simultaneously recover 3D point coordinates and point-to-edge distances from upsampled features and an edge-aware joint loss function to directly minimize distances from output points to 3D meshes and edges. 
However, EC-Net cannot capture line-type features very precisely. 
Loizou et al.~\shortcite{loizou2020learning} formulated the detection of sharp edges as a classification problem and adopted the extended EdgeConv~\cite{dgcnn} to accomplish this task. \SQ{It has to leverage an additional post-processing step of Graph-Cut~\cite{boykov2001fast} to generate reliable results.}
Wang et al.~\shortcite{wang2020pie} 
suggested representing edges as a collection of parametric curves and proposed an end-to-end learnable technique to identify feature edges.
PC2WF~\cite{pc2wf} introduced an end-to-end trainable deep network architecture to accomplish this challenging task, i.e., directly converting a 3D point cloud into a wireframe model. \SQ{Nevertheless, the performance of these data-driven approaches could be influenced by the quantity, quality and relevance of the training dataset.}
\XR{Such approaches~\cite{PIE_Net} may produce promising results for a similar input but could be inadequate for different conditions (e.g., noise level).}

\paragraph{Feature Preserving Surface Reconstruction}
Sharp features need to be carefully handled 
in many geometry processing tasks. 
A typical situation is that the target surface
may consist of smooth patches separated by feature lines~\cite{huang2008surface, zhang2015guided, shen2022gcn}, especially in the field of CAD.
Owing to the piecewise shape representation, 
Du et al.~\shortcite{du2021boundary} 
proposed a representation of Boundary-Sampled Halfspaces (BSH). 
Therefore, it is necessary 
to preserve line-type features 
in reconstructing a CAD model.
{\"O}ztireli et al.~\shortcite{RIMLS} proposed to detect sharp features automatically by measuring the differences between normal vectors. However, the final surface is still differentiable (and thus smooth).
Wang et al.~\shortcite{wang2013feature} 
gave a kernel-based scale estimator
that is used to estimate the best tangent planes and remove outliers.
Salman et al.~\shortcite{salman2010feature}
suggested using a feature detection process based on the covariance matrices of Voronoi cells to extract a set of sharp features,
which facilitates feature preserving mesh generation.
Dey et al.~\shortcite{dey2012feature} 
proposed to identify and reconstruct feature curves
based on the  combination of the Gaussian weighted graph Laplacian and the Reeb graphs. The surface reconstruction step of their algorithm
is akin to the Cocone reconstruction~\cite{Cocone} 
except that the algorithm uses a weighted
Delaunay triangulation technique that allows protection of the feature samples with balls.
Digne et al.~\shortcite{digne2014feature} advocated for iteratively simplifying the initial 3D Delaunay triangulation through optimal transport, where sharp features and boundaries are well-preserved due to the usage of a feature-sensitive metric
between point sets and triangle meshes. Xiong et al.~\shortcite{xiong2014robust} 
proposed a framework that optimizes mesh geometry and connectivity simultaneously from the unoriented point cloud based on dictionary learning~\cite{wright2010sparse}. 
\SQ{Empirical evidence shows that the task of feature preserving reconstruction 
is challenging when the points in the edge zone are sparse.}

%% file: Method.tex
\section{Method}
\label{sec:method}
Given a noisy point cloud of a CAD-type model,
the goal is to reconstruct a clean polygonal surface with neat line-type features. 
\SQ{The most important task is to predict additional points that are able to encode the edge of the geometry.
For that, it is necessary to regularize point locations and normal vectors in advance.
This inspires us to propose a multi-stage algorithm;} See Figure~\ref{FIG:pipeline}. 

\begin{enumerate}
\setlength{\itemindent}{1em}
\item [\bf{Step 1.}] Initialize the normal vectors and 
filter out the noise of the input point cloud. The two operations are conducted at the same time; See Figure~\ref{FIG:pipeline}(b).
\item [\bf{Step 2.}] Identify the points in the edge zone by discrete optimal transport; See Figure~\ref{FIG:pipeline}(c).
\item [\bf{Step 3.}] Refine normal vectors based on the assumption that the target model is locally planar; See Figure~\ref{FIG:pipeline}(d). 
\item [\bf{Step 4.}] Fine-tune point locations to adapt to the regularized normal information; See Figure~\ref{FIG:pipeline}(e).
\item [\bf{Step 5.}] Predict additional points that are on the potential edges of the geometry; See Figure~\ref{FIG:pipeline}(f). 
\item [\bf{Step 6.}] Compute the restricted power diagram (RPD)
on the base surface produced by screened Poisson reconstruction (SPR); See Figure~\ref{FIG:pipeline}(g).
\item [\bf{Step 7.}] Extract the dual of the RPD, giving the reconstructed polygonal surface that recovers the missing feature lines; See Figure~\ref{FIG:pipeline}(h). 
\end{enumerate}

\SQ{In the following, we explain why the steps are organized in such a style. 
Step~1 and Step~2 are used to find the edge zone, which does not depend on highly accurate point locations or normal vectors.
Step~3, Step~4, and Step~5 are used to generate additional points  on the edges of the geometry. 
The underlying observation is that a sufficiently small area around an edge point can be approximated by two half-planes (a feature-line point) or more half-planes (a feature-tip point).
Step~6 and Step~7 focus on interpolating the augmented point set with a triangle mesh, where the key is to guarantee that the connections between predicted edge points exactly report the feature line.}


\subsection{Point Cloud Denoising and Normal Initialization}
\label{sec:init}

Denoising the given noisy point cloud $P=\left \{ p_i \right \}_{i=1}^n$
is a necessary step to guarantee that the final reconstructed result consists mainly of simple patches. 
As point locations $\left \{ p_i \right \}_{i=1}^n$ and normal vectors $\left \{ \mathbf{n}_i \right \}_{i=1}^n$ 
are mutually influenced,
we optimize them jointly. 
By allowing a point $p_i$ to move along the normal direction
$\mathbf{n}_i$, we use
$p_i' = p_i + \epsilon_i \mathbf{n_i}$ 
to denote the new position of $p_i$.
Let
\begin{equation}
M_{3\times 3}^{i}\triangleq
\sum_{p_j\in\text{Neigh}(p_i)}(p_i' - p_j')(p_i' - p_j')^T,\end{equation}
where the neighborhood of $p_i$ 
is measured by a $r$-radius ball. 
In the default setting, $r=2\delta$, i.e., twice the average gap~$\delta$ between points\footnote{
In this paper, we compute~$\delta$ as follows.
First, we find the nearest six neighbors for each point and keep down six distance values.
Second, we compute~$\delta$ by averaging the $6n$
distance values, where $n$ is the number of vertices.
}.
$M_{3\times 3}^{i}$ 
is the covariance matrix, at $p_i$, whose eigenvectors
encode the orientation of the local surface.
If $\mathbf{n_i}$ can reflect the real normal information at $p_i$,
then $M_{3\times 3}^{i}\mathbf{n}_i$ should be close to a null vector.
At the same time, we have to control
the degree of denoising. 
\SQ{Inspired by Principal Component Analysis (PCA)~\cite{pearson1901liii},} we have the following objective function with $\{\epsilon_i\}, \{\mathbf{n}_i\}$ being the joint variables:
\begin{equation}
\centering
\begin{aligned}
\min_{\{\epsilon_i\}, \{\mathbf{n}_i\}} \left\{\sum_{i=1}^n \left\|M_{3\times 3}^{i} \mathbf{n}_i\right\|^2 +\xi\sum_{i=1}^n\epsilon_i^2\right\},
\end{aligned}
\label{eq:denoise_init}
\end{equation}
where $\xi$ is a parameter to balance noise removal and fidelity;
We set $\xi=0.1$ by default in our experiments.
Figure~\ref{FIG:pipeline}(b) shows the denoised result.


\paragraph{Optimization details}
\SQ{
To guarantee $\mathbf{n}_i$
to be a unit vector, 
we parameterize it as
$$\left(\sin(u)\cos(v),\sin(u)\sin(v),\cos(u)\right).$$
Initially, $\epsilon_i=0, i=1,2,\cdots,n.$ 
We initialize $\{\mathbf{n}_i\}$ by the PCA implementation included in PCL~\cite{Rusu_ICRA2011_PCL}. 
Note that it takes orientation consistency into account.
We test how our algorithm is sensitive to the orientation consistency in Section~\ref{sec:limitation} by randomly flipping 5\% to 20\% of the total orientations.
}

\paragraph{Remark}
\SQ{
We observe that the step of denoising 
and the step of detecting the edge zone 
are mutually influenced, i.e.,
it is impossible
to completely eliminate data noise before the edge zone is detected.
Therefore, Eq.~(\ref{eq:denoise_init})
makes the distribution of points better comply with the features of a CAD model, thus enhancing the robustness of the subsequent steps.
It is not purely for denoising points and normal vectors. 
}

\subsection{Edge Zone Identification}
\label{sec:zone}

\begin{figure}[!t]
\centering
\vspace{-1mm}
\begin{overpic}
[width=.85\linewidth]{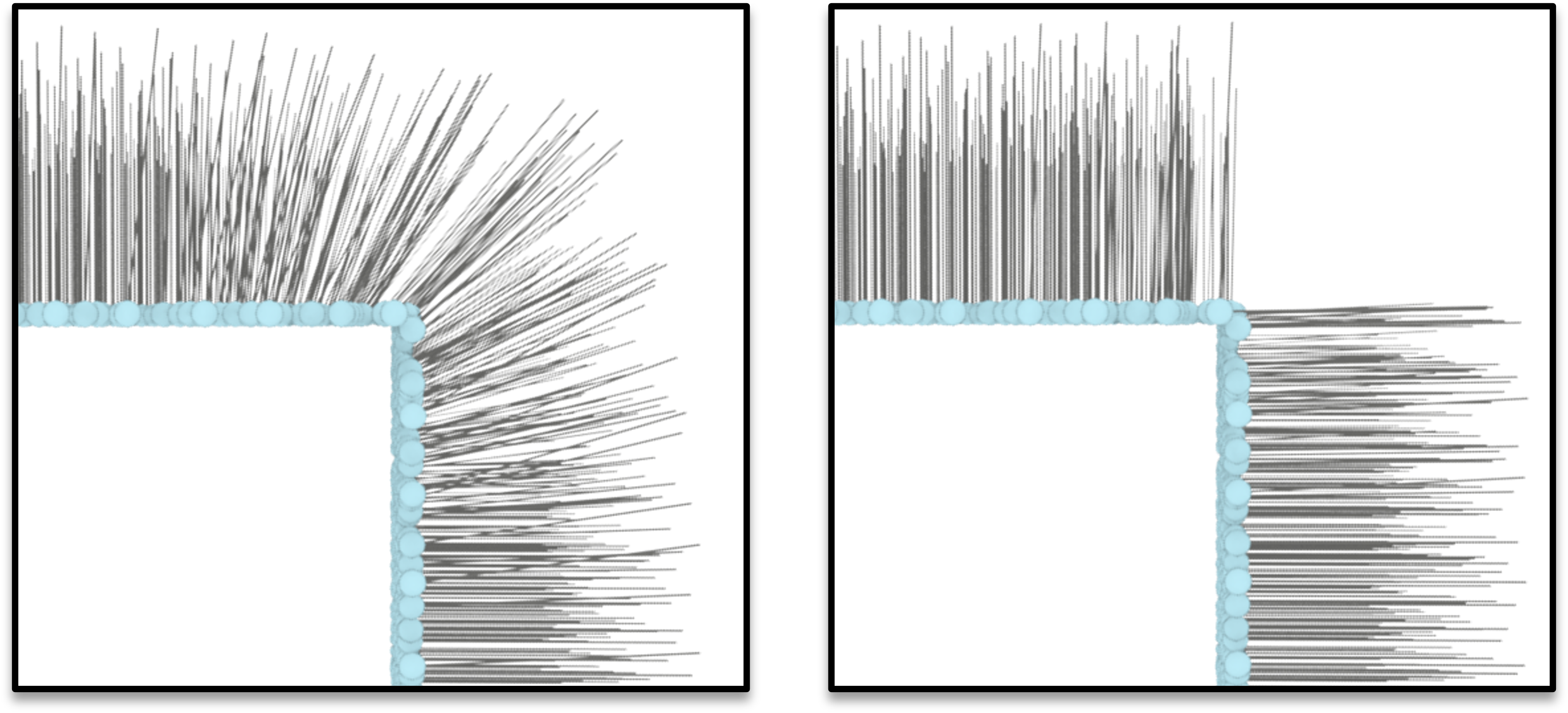}
\put(23,-3){(a)}
\put(76,-3){(b)}
\end{overpic}
\caption{\SQ{An illustrative
example for demonstrating the source distribution~(a) and the target distribution~(b).}}
\vspace{-4mm}
\label{FIG:normal}
\end{figure}

\paragraph{Observation}
Generally, the surface of a CAD model consists of multiple smooth-and-simple pieces.  
Suppose that the point $p_i$ is located in a nearly flat region. 
The probability density function (PDF) of the normal vectors nearby $p_i$ can be approximated by a single Dirac delta function (multiplied by a vector). If $p_i$ is very close to an edge, 
instead, the PDF can be represented as a combination of two separate Dirac delta functions. Furthermore, the two sub-distributions have nearly equal quantities. (If $p_i$ is near a corner point, the PDF can be decomposed into three or more separate Dirac delta functions.) 
\SQ{Suppose that $p_i$ is located on the edge of the geometry; See Figure~\ref{FIG:normal}. 
The yielded normal vectors by Step~1 are visualized in Figure~\ref{FIG:normal}(a), and they define the source distribution. 
The target distribution is given by two separate normal vectors (shown in Figure~\ref{FIG:normal}(b)).
Rather than use k-means to explicitly cluster the normal vectors into two groups,
we provide a more accurate measure to characterize the difference between the source distribution and the target distribution.}

\paragraph{Optimal Mass Transport.}
\XR{Optimal Mass Transport (OMT) \cite{monge1781memoire} aims at finding the lowest transportation cost of moving the source distribution $\mu_s$ to the target distribution $\mu_t$, where both the source domain and the target domain are given by the surface of the unit sphere $S^2$.
\begin{equation}
\begin{aligned}
\inf_{f}
\int_{x\in S^2} c(x,f(x))d\mu_s,
\end{aligned}
\end{equation}
where $f:S^2\rightarrow S^2$ is a $1-1$ mapping, and $c(x,f(x))$ describes the cost for transporting a unit mass from $x\in S^2$ to $f(x)\in S^2$. In this paper, $c(\cdot,\cdot)$ is simply estimated as the squared difference between two unit normal vectors; See Eq.~(\ref{eq:feature_point}).
In our scenario, we define the source distribution $\mu_s$ by a set of normal vectors produced in Sec.~\ref{sec:init} and the target distribution $\mu_t$ as two independent Dirac delta functions; See Figure~\ref{FIG:normal}(b). 
}


\begin{figure}[!t]
	\centering
\begin{overpic}
[width=.85\linewidth]{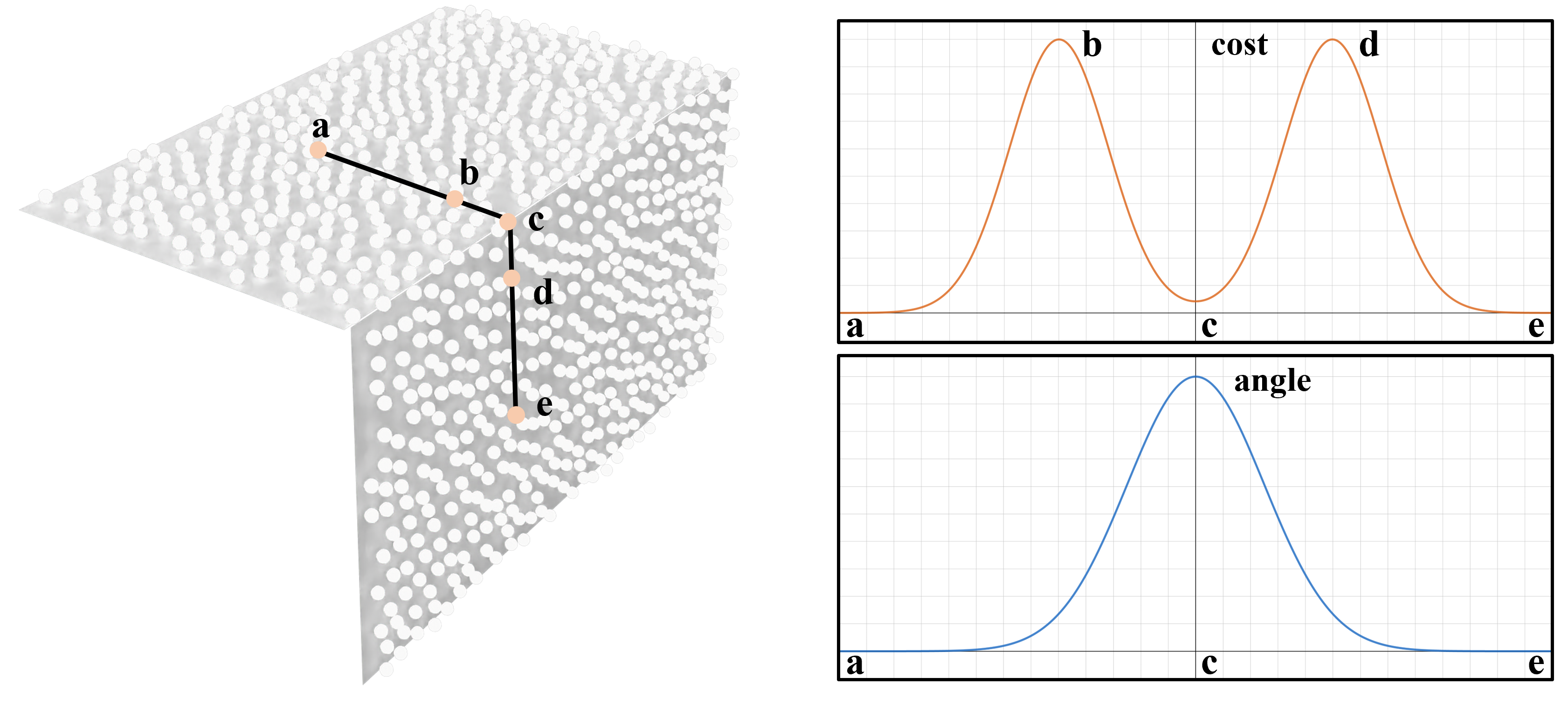}
\end{overpic}
\vspace{-3mm}
\caption{We plot the transport cost given by Eq.~(\ref{equa:FeatureZone}) 
and the angle between $\hat{\mathbf{n}}_1$ and $\hat{\mathbf{n}}_2$
by moving a point across the edge of the geometry. 
}
\vspace{-4mm}
\label{FIG:cost}
\end{figure}

\paragraph{Formulation.} 
We consider the situation when $p_i$ is located on a geometry edge.
It can be observed that in the neighborhood of~$p_i$,
the normal vectors can be organized into two equal-quantity clusters whose representative normal vectors are  $\hat{\mathbf{n}}_1$ and $\hat{\mathbf{n}}_2$, respectively.
Therefore, the squared transport cost
\begin{equation}
\centering
\min_{\{\lambda_j\}}\sum_{p_j\in\text{Neigh}(p_i)}\left\{\lambda_j\left \| \mathbf{n}_j-\hat{\mathbf{n}}_1 \right \|^2
+(1-\lambda_j)\left \| \mathbf{n}_j-\hat{\mathbf{n}}_2 \right \|^2\right\},
\label{eq:feature_point}
\end{equation}
is very close to 0, where $\lambda_j\in[0,1]$ is used to define the proportion of $\mathbf{n}_j$ transported to $\hat{\mathbf{n}}_1$.
When the above objective function gets minimized,
\begin{equation}
\lambda_j=
  \begin{cases}
   1        & \text{$\mathbf{n}_j$ is closer to $\hat{\mathbf{n}}_1$}, \\
   0        & \text{otherwise}.
  \end{cases}
\end{equation}
Since the two clusters of normal vectors have the nearly equal quantity, we have
\begin{equation}
    \sum_{p_j\in\text{Neigh}(p_i)}\lambda_j\approx\frac{k}{2},
\end{equation}
where $k$ is the number of points in $p_i$'s neighborhood. 
This inspires us to characterize the edge zone by the following formulation:

\begin{equation}
\label{equa:FeatureZone}
\begin{aligned}
&\min_{\{\lambda_j\}_{j=1}^k, \hat{\mathbf{n}}_1, \hat{\mathbf{n}}_2} 
\sum_{p_j\in\text{Neigh}(p_i)}\left\{\lambda_j \rho_{1}
+(1-\lambda_j)\rho_{2}\right\},\\
&\text { s.t. } \begin{cases}
\rho_{d}=\left\|\mathbf{n}_{j}-\hat{\mathbf{n}}_{d}\right\|^{2}, & d=1,2, \\
0\leq\lambda_j\leq 1, &j=1,2,3,\ldots,k,\\
 \sum_{j=1}^k \lambda_j=\frac{k}{2}, \\
\|\hat{\mathbf{n}}_1\|=1,  \|\hat{\mathbf{n}}_2\|=1.\end{cases}
\end{aligned}
\end{equation}
The above formulation 
is to characterize the degree of deviation from
the current vector-normal distribution
to a ``perfect'' geometry edge (\SQ{the normal vector for a point in the neighborhood is either $\hat{\mathbf{n}}_1$ or $\hat{\mathbf{n}}_2$; the number of points with $\hat{\mathbf{n}}_1$ is equal to that with $\hat{\mathbf{n}}_2$}). 
To better explain the insight, we make a toy model, as Figure~\ref{FIG:cost} shows,
to observe how the transport cost of Eq.~(\ref{equa:FeatureZone})
changes across the edge of the geometry.
On the one hand, it can be seen that the transport cost 
is very close to 0 for a point in the flat region or on the edge. 
On the other hand, 
the angle between $\hat{\mathbf{n}}_1$ and $\hat{\mathbf{n}}_2$
reaches the maximum when the point arrives at the edge (like point $c$ in Figure~\ref{FIG:cost}). 
We propose a geometrically meaningful rule to identify 
the edge zone. 

\paragraph{Three situations.}
\SQ{
Upon Eq.~(\ref{equa:FeatureZone}) being optimized,
the angle between $\hat{\mathbf{n}}_1$ and $\hat{\mathbf{n}}_2$,
denoted by $\text{Angle}(\hat{\mathbf{n}}_1,\hat{\mathbf{n}}_2)$,
can be used to distinguish flat regions from edge zones. 
But we observe that if $p_i$ is located on thin tubes,
both the angle $\text{Angle}(\hat{\mathbf{n}}_1,\hat{\mathbf{n}}_2)$ and the transport cost (the function value) $\text{Cost}(p_i)$ are large. Therefore,
we consider the following three situations:\\
\indent { {\bf Situation 1:} $\text{Angle}(\hat{\mathbf{n}}_1,\hat{\mathbf{n}}_2)\leq \pi/6$. We take the base point $p_i$ as an off-edge point.}\\
\indent{ {\bf Situation 2:} $\text{Angle}(\hat{\mathbf{n}}_1,\hat{\mathbf{n}}_2)> \pi/6$
and $\text{Cost}(p_i)\leq 0.25$.
We 
label the base point $p_i$ with ``edge-zone''.} \\
\indent{ {\bf Situation 3:} 
$\text{Angle}(\hat{\mathbf{n}}_1,\hat{\mathbf{n}}_2)> \pi/6$
and $\text{Cost}(p_i)>0.25$.
We also take $p_i$ as an off-edge point. } 
}

\paragraph{Implementation Details.}
\XR{It's worth noting that the step of edge zone identification 
assumes that the problem of normal orientation consistency for the whole point cloud has been addressed 
in the previous step.}
\SQ{
The initial values of~$\hat{\mathbf{n}}_1$
and $\hat{\mathbf{n}}_2$ are given by running k-means for the~$k$ normal vectors. Each coefficient~$\lambda_j$ is initialized to 0.5.
At the same time, we introduce an adaptive weighting scheme in Eq.~(\ref{equa:FeatureZone}) to deal with uneven point distribution. }
\begin{equation}
\label{eq:edge_zone}
\centering
\sum_{p_j\in\text{Neigh}(p_i)}
\frac{\lambda_j\left \| \mathbf{n}_j-\hat{\mathbf{n}}_1 \right \|^2
+(1-\lambda_j)\left \| \mathbf{n}_j-\hat{\mathbf{n}}_2 \right \|^2}{\left \| p_i-p_j \right \|^2 +\epsilon},
\end{equation}
where $\epsilon$ ($10^{-4}$ by default) is used to deal with the existence of a vanishing denominator. 


\subsection{Edge Point Prediction} 
\label{sec:FeaturePoint}
\SQ{To this end,
we come to fine-tune normal vectors and point locations
according to the detected edge zone, and then generate additional edge points. }

\emph{Regularizing Normal Vectors.}
After the edge points are detected, 
their normal vectors need to be re-computed as the normal vectors have an abrupt change across the edges of the geometry. 
Let $\mathcal{F}$ be the edge-point set. 
For each point $p_i\in \mathcal{F}$, there are two typical cases:\\
\indent{\bf Case 1:} $p_i$ is close to a feature-tip point, and
the normal vectors given by $p_i$'s neighborhood 
have at least three clusters.\\
\indent{\bf Case 2:} $p_i$ is close to a feature-line point
but far from a feature-tip point,
and
the normal vectors given by $p_i$'s neighborhood 
have two clusters.

\begin{figure}[!t]
	\centering
\begin{overpic}
[width=\linewidth]{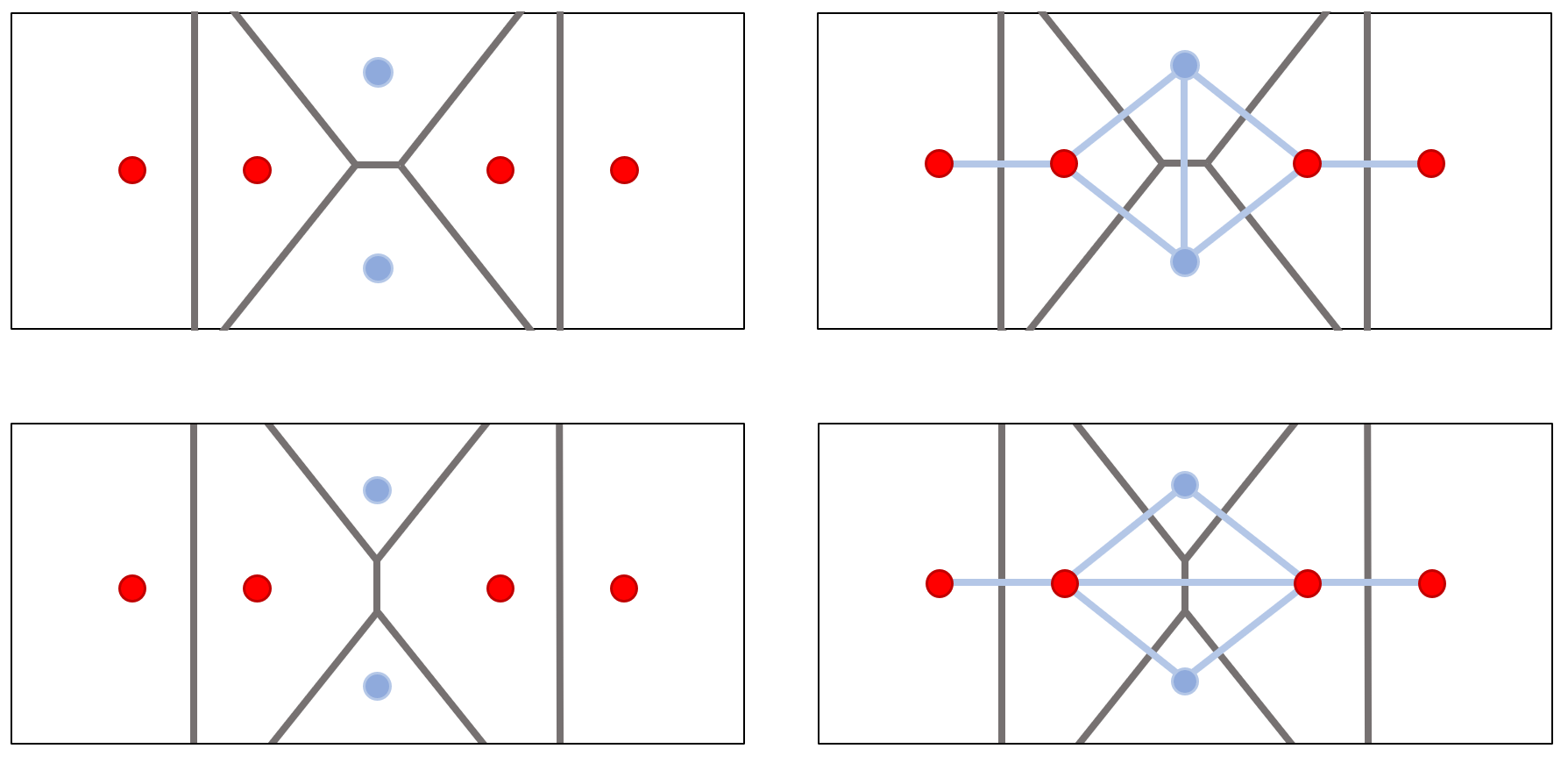}
\put(8,24){(a)~Voronoi diagram}
\put(64,24){(b)~Dual of~(a)}
\put(10,-2){(c)~Power diagram}
\put(64,-2){(d)~Dual of~(c)}
\end{overpic}
\caption{
A 2D example for explaining why the power diagram helps preserve feature lines, where the points colored in red can be viewed as edge points.
(a) The Voronoi diagram takes each site with equal importance.
(b) The resulting triangulation (dual of (a)) fails to align the connections with the potential feature line due to the insufficient density of edge points.
(c) By using the power diagram, one increases the influence of edge points. 
(d) The edge points are given higher priority when inferring the connections between points, making the resulting triangulation (dual of (c)) feature-line aligned. }
	\label{FIG:power}
\vspace{-2mm}
\end{figure}
\begin{figure}[!t]
\vspace{-3.0mm}
	\centering
\begin{overpic}
[width=\linewidth]{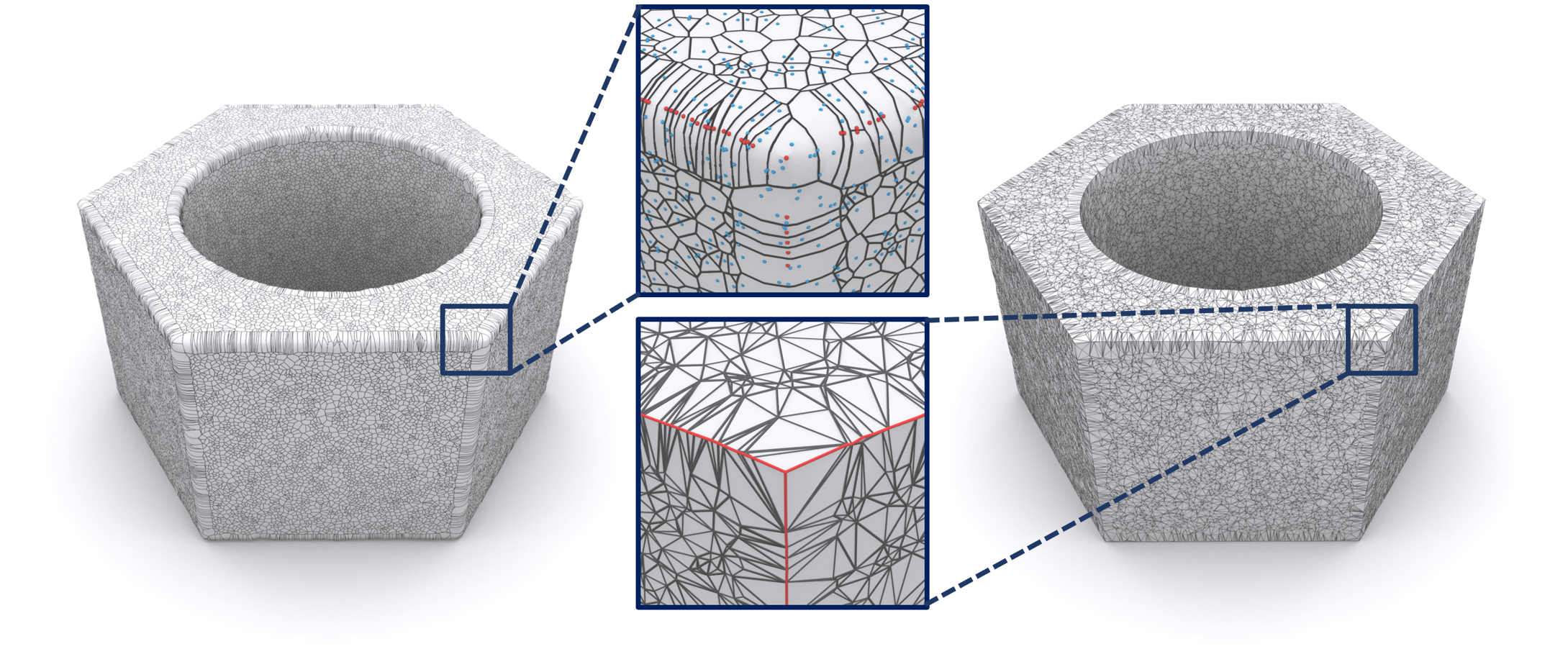}
\put(0,-1.5){(a)~Restricted power diagram}
\put(63,-1.5){(b)~Dual triangulation}
\end{overpic}
\caption{The RPD-based surface reconstruction. (a)~The RPD on the Screened Poisson reconstruction surface. (b)~Connect the augmented point set into a triangle mesh based on the dual of the RPD.
}
\vspace{-5mm}
\label{FIG:rvd}
\end{figure}

\SQ{To fine-tune the normal vector at~$p_i$,
we follow Eq.~(\ref{equa:FeatureZone}) 
but relax the requirement of quantity equilibrium,
achieving the following optimization for regularizing normal vectors. }
\begin{equation}
\label{eq:three}
\begin{aligned}
&\min _{\left\{\lambda_{j}^{(1,2,3)}\right\}_{j=1}^{k}, \hat{\mathbf{n}}_{1}, \hat{\mathbf{n}}_{2},\hat{\mathbf{n}}_{3}}
\sum_{p_j\in\text{Neigh}(p_i)} \frac{\lambda_{j}^{(1)} \rho_{1}+\lambda_{j}^{(2)} \rho_{2}+\lambda_{j}^{(3)} \rho_{3}}{\left\|p_{i}-p_{j}\right\|^{2}+\epsilon}, \\
&\text { s.t. } \begin{cases}\rho_{d}=\left\|\mathbf{n}_{j}-\hat{\mathbf{n}}_{d}\right\|^{2}, & d=1,2,3, \\
\lambda_{j}^{(1,2,3)}=0, & p_{j} \in \mathcal{F}, \\
0 \leq \lambda_{j}^{(1,2,3)} \leq 1, & p_{j} \notin  \mathcal{F}, \\
\lambda_{j}^{(1)}+\lambda_{j}^{(2)}+\lambda_{j}^{(3)}=1, & p_{j} \notin  \mathcal{F}, \\
\left\|\hat{\mathbf{n}}_{1}\right\|=1,\left\|\hat{\mathbf{n}}_{2}\right\|=1,\left\|\hat{\mathbf{n}}_{3}\right\|=1. & \end{cases}
\end{aligned}
\end{equation}
\XR{
Obviously, Eq.~(\ref{eq:three}) depends on normal orientation consistency.
We assume that the consistency problem is addressed 
in Step~1.}
At the end of the optimization,
we compute $\sum_{j=1}^k\lambda_{j}^{(1)}$,
$\sum_{j=1}^k\lambda_{j}^{(2)}$,
and $\sum_{j=1}^k\lambda_{j}^{(3)}$ to see which is the largest. 
We then use the most significant normal vector among $\hat{\mathbf{n}}_{1}$, $\hat{\mathbf{n}}_{2}$,
and $\hat{\mathbf{n}}_{3}$ to update $p_i$'s normal vector.
\SQ{The whole regularization step is accomplished by repeating the operation for each point;
See Figure~\ref{FIG:pipeline}(d) for illustration.
It's worth noting that 
both Eq.~(\ref{eq:edge_zone}) and Eq.~(\ref{eq:three})
are operated by optimization, instead of k-means. 
We discuss the difference between k-means and our formulation in Section~\ref{sec:kmeans_init}.
} 


\emph{Point Location Refinement.} 
Next we come to jointly optimize the point locations  to coincide with the regularized normal vectors. 
Like the formulation of Eq.~(\ref{eq:denoise_init}),
we suggest the following optimization:
\begin{equation}
\centering
\begin{aligned}
&\min_{\{\epsilon_i\}} 
\sum_{i=1}^n\sum_{p_j\in \text{Neigh}'(p_i)}\left\|M_{3\times 3}^{ij} \mathbf{n}_i\right\|^2,\\
\end{aligned}
\label{eq:refinement}
\end{equation}
where $\text{Neigh}'(p_i)\subset \text{Neigh}(p_i)$ includes 
only those neighbors with similar normal vectors, i.e., $\langle \mathbf{n}_j,\mathbf{n}_i \rangle  \leq \frac{\pi}{6}$.
Furthermore, it is worth noting that Eq.~(\ref{eq:refinement})
does not include the term $\sum \epsilon_i^2$.
The normal vectors, after being regularized, are more faithful, and thus excluding the term can make the resulting point cloud more locally planar. 
See Figure~\ref{FIG:pipeline}(e) for the refinement result. 

\emph{Edge Point Generation.}
Let $p_i$ be a point  identified in the edge zone.
Each point $p_j\in\text{Neigh}(p_i)$ defines a tangent plane $\Pi\triangleq(p_j,\mathbf{n}_j)$, and $\Pi$ goes through the nearby geometry edge. 
\SQ{Similar to \cite{chen2021multiscale}, we project~$p_i$ to the closest point on a feature line by introducing following optimization with long-distance punishment $\left \| z_i - p_i \right \|^2$:}
\begin{equation}
    \mathop{\min}_{z_i} \sum_{p_j\in\text{Neigh}(p_i)}\left((z_i-p_j)\cdot \mathbf{n}_j)\right)^2 + \mu \left \| z_i - p_i \right \|^2,
    \label{equa:6}
\end{equation}

where $\mu$ helps pull the projection $z_i$ to the closest point on the feature line. 
(Note that, if $p_i$ is near a corner point, 
minimizing $\sum\left((z_i-p_j)\cdot \mathbf{n}_j)\right)^2$ can also pull $p_i$ towards the corners.)
By default, $\mu$ is set to 0.01 in our experiments. 
See Figure~\ref{FIG:pipeline}(f) for generated edge points.

\begin{figure}[!t]
\vspace{-2mm}
	\centering
\begin{overpic}
[width=0.80\linewidth]{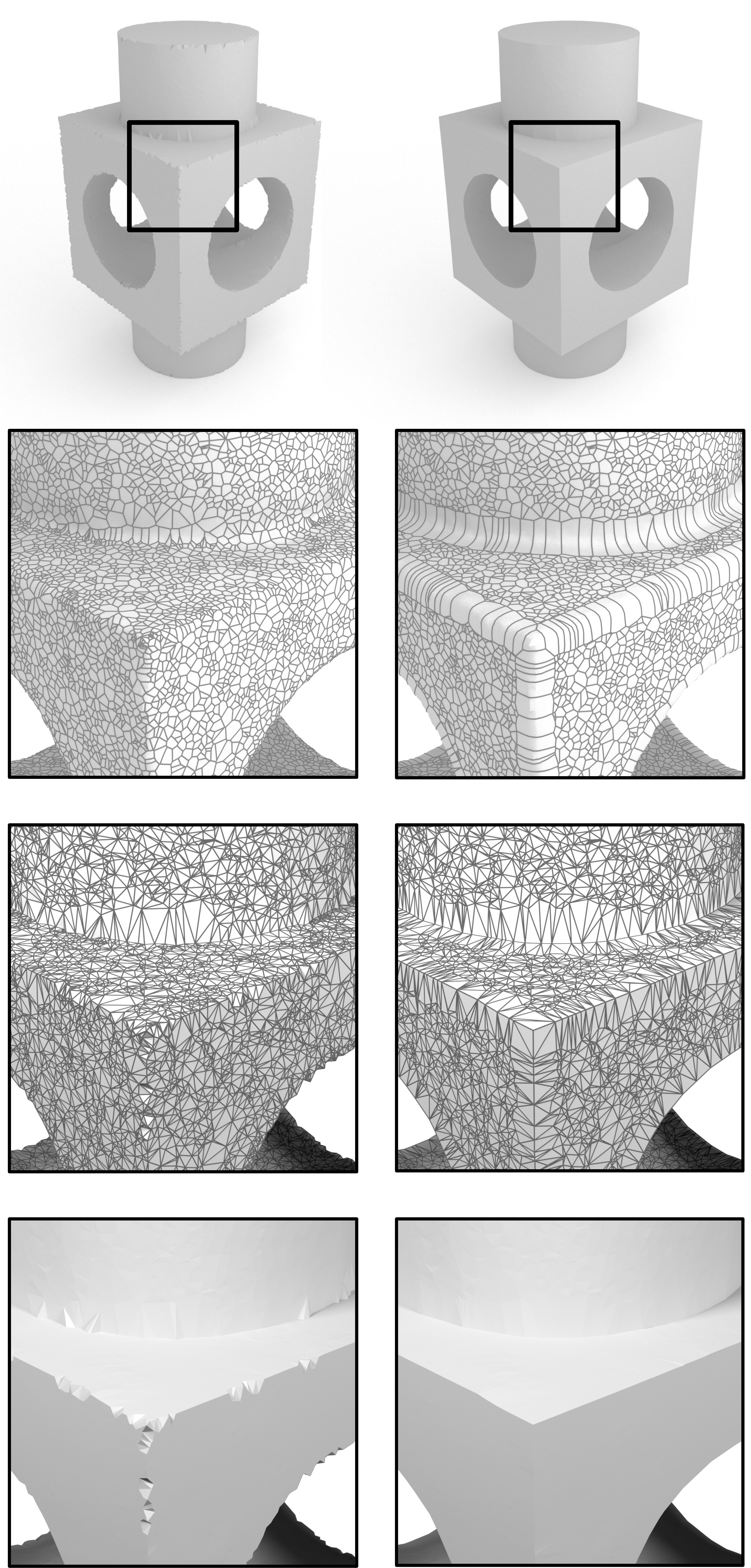}
\put(4,73.5){(a)~Result of RVD}
\put(28,73.5){(b)~Result of RPD}
\put(-4.5,48.5){(c)~Restricted Voronoi Diagram}
\put(24,48.5){(d)~Restricted Power Diagram}
\put(4,23){(e)~Dual of RVD}
\put(28,23){(f)~Dual of RPD}
\put(4,-2){(g)~Dual of RVD}
\put(28,-2){(h)~Dual of RPD}
\end{overpic}
\caption{The close-up window shows that the RPD is better at preserving feature lines than the RVD for surface reconstruction. 
}
\label{FIG:rvdrpd}
\vspace{-4mm}
\end{figure}
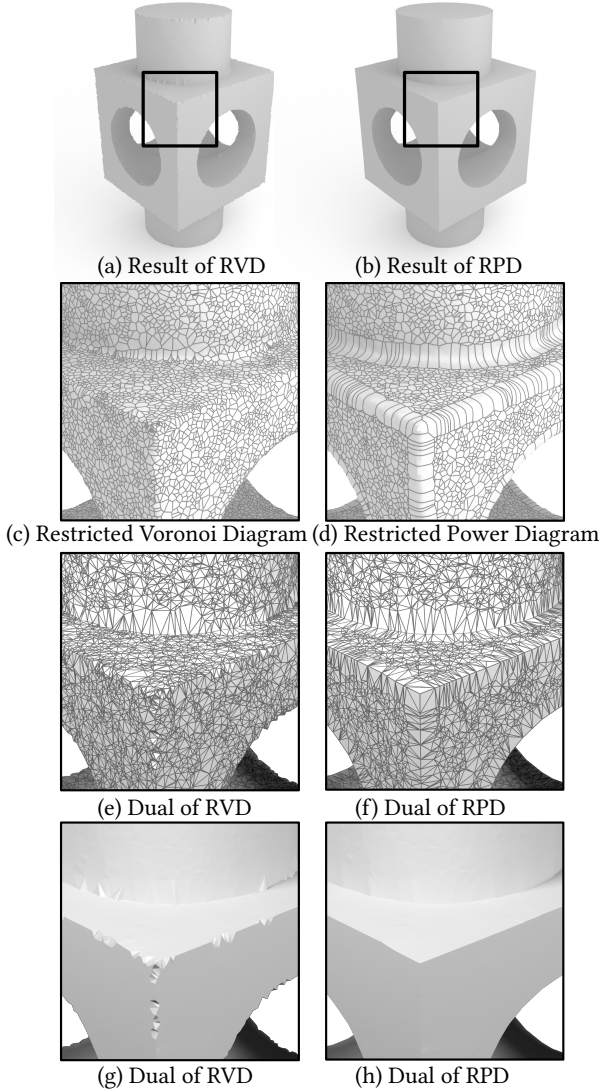
\subsection{Feature Preserving Interpolation-Based Reconstruction}
\label{subsec:RPD_recon}
The final stage involves surface reconstruction. 
As it is hard for implicit approaches to  handle feature lines, 
we generate a polygonal surface that interpolates the augmented point (edge points included). 
The restricted Voronoi diagram~(RVD)~\cite{edelsbrunner1994triangulating,yan2009isotropic,yan2014low} is a commonly used tool for mesh generation. 
As Khoury and Shewchuk~\shortcite{khoury2016fixed} pointed out, 
the RVD can yield a high-quality approximation of the underlying surface if one can find a base surface that is sufficiently close to the target surface (the admissible deviation is related to local feature size). 
However, the Voronoi Diagram treats each site with equal importance, thus the resulting triangulation will not be explicitly aligned with the potential feature line if the edge points are not dense enough; See Figure~\ref{FIG:power}(a,b).
We observe that the power diagram is a better tool for treating this problem, where the edge points can be given a larger influence; See Figure~\ref{FIG:power}(c,d).

\begin{figure*}[!t]
	\centering
\begin{overpic}
[width=.99\textwidth]{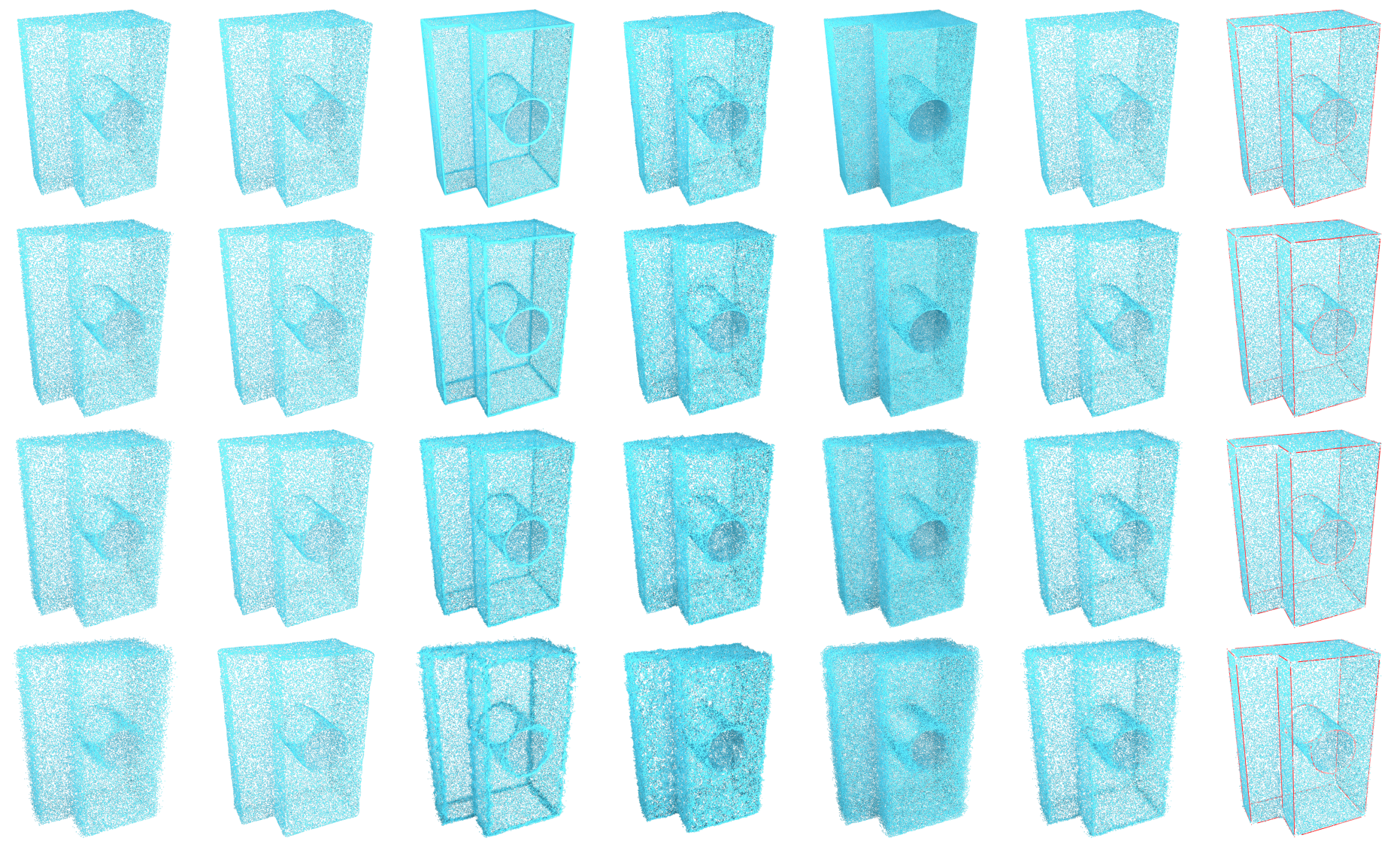}
\put(5,-1.5){\textbf{Input}}
\put(20,-1.5){\textbf{RIMLS}}
\put(34,-1.5){\textbf{EAR}}
\put(47.5,-1.5){\textbf{EC-Net}}
\put(61.5,-1.5){\textbf{Dis-PU}}
\put(76.5,-1.5){\textbf{MFLE}}
\put(91.5,-1.5){\textbf{Ours}}
\put(-0.5,48){\rotatebox{90}{\textbf{no-noise}}}
\put(-0.5,33){\rotatebox{90}{\textbf{0.25\%-noise }}}
\put(-0.5,18){\rotatebox{90}{\textbf{0.5\%-noise }}}
\put(-0.5,4){\rotatebox{90}{\textbf{1\%-noise }}}
\end{overpic}
\caption{Test point cloud consolidation approaches by introducing different levels of noise. It can be seen that our method can not only effectively eliminate noise but also recover faithful feature lines. Statistics can be found in Table~\ref{tab:quantitative}.}
	\label{FIG:Consolidation}
\vspace{-4mm}
\end{figure*}

In our scenario, we first generate the base surface by running the screened Poisson reconstruction solver on the augmented point set.
\SQ{Next, we project each point onto the reconstructed surface
and use the restricted power diagram~(RPD)~\cite{basselin2021restricted} to infer the connections between points.
Finally, we copy the connections back to the original points, accomplishing the reconstruction task; See Figure~\ref{FIG:rvd}.}
Let $\delta$ be the average gap between points.
In our implementation, the edge points are assigned a weight of $8\delta^2$ while the other points are given a zero weight. 
We further compare RVD and RPD in Figure~\ref{FIG:rvdrpd}.
\SQ{It can be seen that
the RVD does not give special treatment to edge points, and thus the resulting triangulation does not own the property of feature-line alignment. 
In contrast, the RPD is better at keeping points on different sides of a geometry edge disconnected. }

%% file: Results.tex
\begin{table}[!htp]
\centering
\caption{\SQ{Evaluating various point consolidation approaches for point cloud inputs with varying levels of noise.}}
\label{tab:quantitative}
\vspace{-3mm}
\resizebox{0.475\textwidth}{!}{
\begin{tabular}{l|cccc|cccc} 
\toprule
       & \multicolumn{4}{c|}{$\mathrm{OCD}\left(\times 10^{4}\right) \downarrow$} & \multicolumn{4}{c}{$\mathrm{OECD}\left(\times 10^4\right) \downarrow$}   \\ 
\cmidrule{2-9}
       & no-noise       & 0.25\%         & 0.5\%          & 1.0\%                 & no-noise       & 0.25\%         & 0.5\%          & 1.0\%  \\ 
\cmidrule(l){1-9}
RIMLS  & 0.089          & 0.189          & 0.327          & \textbf{0.544}        & 0.108          & 0.136          & 0.139          & 0.149  \\
EAR    & 0.085          & 0.308          & 0.948          & 3.202                 & 0.097          & 0.140          & 0.142          & 0.147   \\
EC-Net & 0.215          & 0.337          & 0.450          & 1.337                 & 0.132          & 0.134          & 0.137          & 0.144  \\
Dis-PU & 0.097          & 0.303          & 1.012          & 3.385                 & 0.112          & 0.138          & 0.138          & 0.142  \\
MFLE   & 0.087          & 0.307          & 0.958          & 3.243                 & 0.105          & 0.140          & 0.146          & 0.145   \\
Ours   & \textbf{0.084} & \textbf{0.140} & \textbf{0.316} & 0.572                 & \textbf{0.079} & \textbf{0.102} & \textbf{0.122} & \textbf{0.140}   \\
\bottomrule
\end{tabular}
}
\vspace{-4mm}
\end{table}
\section{Experimental Results}\label{sec:Results}

\subsection{Experimental Setting}
\paragraph{Experimental Platform and Parameters}
We experiment on a computer with an AMD Ryzen 9 5950X CPU and 32 GB memory.
Since some learning-based approaches are included for comparison,
we run them on NVIDIA RTX 2080Ti card. 
All the point cloud models are normalized to a range of $[-0.5,0.5]^3$.
We extract 50K surface samples from each model by white noise sampling~\cite{gptoolbox}. 
\SQ{
All the experiments follow the same parameter setting, i.e., 
$\xi = 0.1$, $r = 2\delta$ and $\mu = 0.01$, 
where the meaning of the parameters can be found
in Eq.~(\ref{eq:denoise_init}) and Eq.~(\ref{equa:6}).
We use the LBFGS function in the ALGLIB solver~\cite{alglib} for solving constrained optimization problems defined in Section~\ref{sec:method},
and we set the constraints by the function ``minbleicsetlc''. 
The termination condition for all the optimization problems is set by requiring the gradient norm not to exceed $10^{-4}$, except that
the tolerance for Eq.~(\ref{equa:6}) is set to $10^{-6}$ for higher accuracy.
Besides, we extend the implementation of the restricted Voronoi diagram~\cite{lpcvt} 
to compute the restricted power diagram.}

\paragraph{Datasets}
Our tests are made on both synthetic and raw-scan data,
where the synthetic point cloud data is sampled from the models in the ABC dataset~\cite{Koch_2019_CVPR}.
As some models have defects (e.g., self-penetration),
we select 100 models from the dataset. 
For each model, we sample 50K points with white noise sampling~\cite{gptoolbox}
to get noise-free data.
Additionally, we add different levels of Gaussian noise to generate noisy data, 
i.e., $0.25\%, 0.5\%, 1\%$ of the diagonal length of the bounding box.

\paragraph{Evaluation Metrics}
The topic of this paper is related to both
point consolidation~\cite{huang2009consolidation} and surface reconstruction. 
We use the one-sided Chamfer Distance (OCD) and one-sided Edge Chamfer Distance (OECD) to measure how close the consolidated point cloud is to the ground-truth surface,
where OECD is obtained by computing the average deviation for those points within a distance of less than 0.005 to feature lines.

To evaluate the accuracy of the reconstructed mesh, 
we use three indicators, including Chamfer Distance (CD), F-score (F1), and Normal Consistency (NC). 
We also use Edge Chamfer Distance (ECD) and Edge F-score (EF1) proposed by NMC~\cite{chen2021nmc} to measure the sharpness of the reconstruction mesh. 

\subsection{Point Cloud Consolidation}
\label{sec:point_comparison}

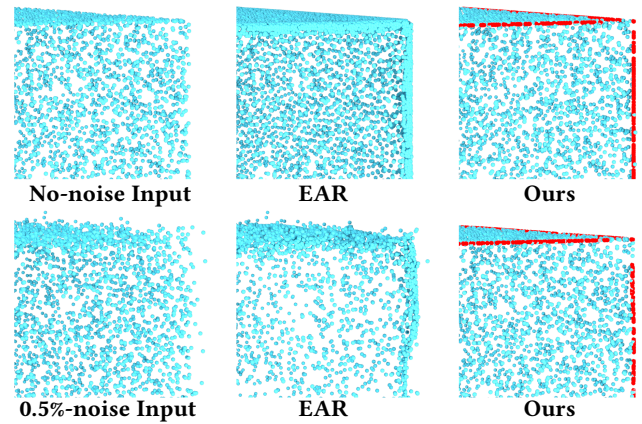
\begin{figure}[!t]
	\centering
\vspace{-2mm}
\begin{overpic}
[width=\linewidth]{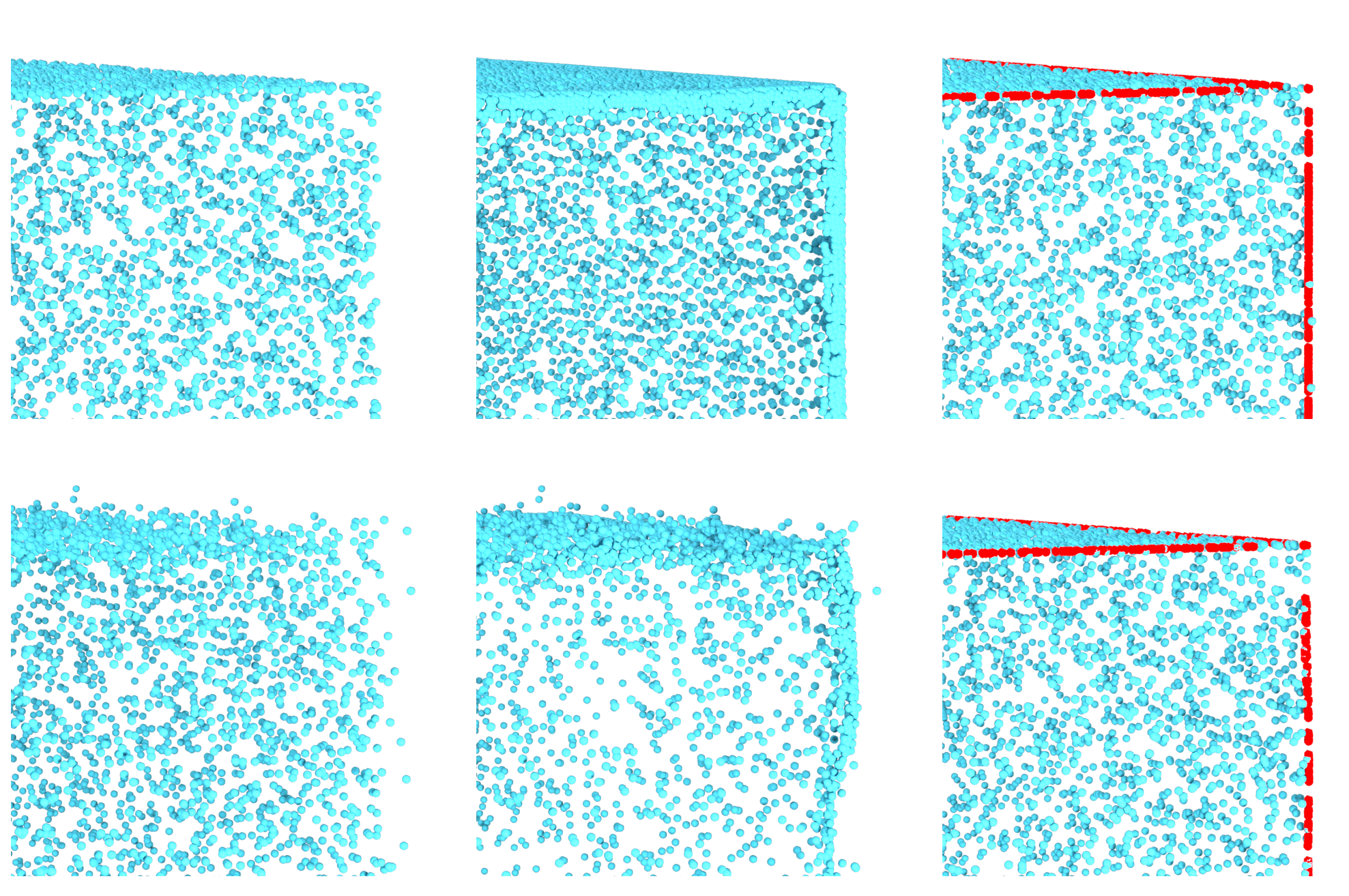}
\put(1.5,-1.5){\textbf{0.5\%-noise Input}}
\put(45,-1.5){\textbf{EAR}}
\put(80,-1.5){\textbf{Ours}}
\put(3,32){\textbf{No-noise Input}}
\put(45,32){\textbf{EAR}}
\put(80,32){\textbf{Ours}}
\end{overpic}
\caption{Test EAR and our approach on clean and noisy data. 
It can be seen that EAR increases the density of the augmented points near the edge region, but the non-regularized distribution does not easily lead to straight-and-smooth feature lines. 
}
\vspace{-5mm}
\label{FIG:CompareEAR}
\end{figure}

We compare our approach with state-of-the-art point cloud consolidation methods, including robust implicit moving least-square (RIMLS)~\cite{RIMLS}, edge-aware resample (EAR)~\cite{EAR}, EC-Net~\cite{yu2018ec}, Dis-PU~\cite{li2021dispu}, and MFLE~\cite{chen2021multiscale}.
At the same time, we introduce $0.25\%$, $0.5\%$, $1\%$ Gaussian noise 
to observe the denoising ability.
Figure~\ref{FIG:Consolidation} gives 
an example of visualizing the difference 
between these point cloud consolidation approaches.
The OCD and OECD statistics are available in    Table~\ref{tab:quantitative}. 
We also give a qualitative 
evaluation in terms of denoising ability, edge awareness, and the normal refinement quality based on our tests. 
For a fair comparison, 
we use the pre-trained models (publicly released) for EC-Net and Dis-PU while tuning the parameters for RIMLS, EAR, and MFLE
so as to achieve the best scores.
It can be seen in Figure~\ref{FIG:Consolidation} 
that the augmented points by our approach better reflect the edge of the geometry. 
The challenge of consolidating a CAD-type point cloud 
lies in accurately inferring the underlying line-type features, but it is hard for RIMLS to reproduce the sharpness of the geometry due to its $C^2$-continuous formulation. 
Similarly, although EAR, EC-Net, and MFLE are also designed to preserve sharp features, 
it is not easy for them to sufficiently reproduce the abrupt change of normal vectors across the geometry edge. 
\SQ{Dis-PU focuses more on the task of upsampling and targets at distribution uniformity and proximity-to-surface, but does not explicitly preserve line-type features.}
We show a close-up result in Figure~\ref{FIG:CompareEAR} to visualize
the difference between EAR and our approach. 
EAR tends to increase the density of the augmented points near the edge region, \SQ{but the additional points do not accurately align with the potential feature lines, especially in the presence of noise. }

To summarize, CAD-type point clouds are structurally distinct from other models, making the operation of data denoising, edge-zone identification, and point consolidation interdependent.
The key idea of our algorithm is based on a weak prior that
the point cloud is locally planar,
and thus, Step 1, Step 3, and Step 4 of our pipeline focus on regularizing point locations and normal vectors.
When point locations and normal vectors become accurate, the edge-point generation step can produce a set of augmented points that align with the edge of the geometry.

\begin{table*}[!htp]
\centering
\caption{
As the reconstruction of a CAD-type point cloud 
includes a step of point consolidation and a step of mesh generation, we use one point consolidation approach
and one surface reconstruction approach to define a combination,
where ``w.o.'' means ``without point consolidation''. The best scores are highlighted in bold.}
\label{tab:quantitative_noisefree}
\resizebox{1.0\linewidth}{!}{
\begin{tabular}{c|l|cccc|cccc|cccc|cccc|cccc} 
\toprule
\multirow{2}{*}{Noise}        & \multicolumn{1}{c|}{\multirow{2}{*}{Methods}} & \multicolumn{4}{c|}{$\mathrm{CD}\left(\times 10^{4}\right) \downarrow$}      & \multicolumn{4}{c|}{$\mathrm{F1}\uparrow$}                        & \multicolumn{4}{c|}{$\mathrm{NC}\uparrow$}                        & \multicolumn{4}{c|}{$\mathrm{ECD}\left(\times 10^{2}\right)\downarrow$} & \multicolumn{4}{c}{$\mathrm{EF1}\uparrow$}  \\ 
\cmidrule{3-22}
                              & \multicolumn{1}{c|}{}                         & w.o.            & RIMLS          & EAR             & Ours                    & w.o.           & RIMLS          & EAR            & Ours           & w.o.           & RIMLS          & EAR            & Ours           & w.o.   & RIMLS  & EAR    & Ours                                         & w.o.  & RIMLS  & EAR   & Ours               \\ 
\midrule
\multirow{6}{*}{no-noise}     & GD                                            & 0.175           & \textbf{0.180} & 0.276           & 0.173                   & 0.701          & \textbf{0.675} & 0.699          & 0.706          & \textbf{0.988} & 0.987          & \textbf{0.990} & \textbf{0.990} & 0.696  & 2.432  & 1.022  & 0.673                                        & 0.355 & 0.172  & 0.413 & 0.431              \\
                              & RIMLS                                         & 0.721           & 0.727          & 3.653           & 0.725                   & 0.403          & 0.407          & 0.571          & 0.370          & 0.978          & 0.978          & 0.959          & 0.968          & 1.102  & 0.918  & 27.272 & 0.625                                        & 0.110 & 0.127  & 0.002 & 0.163              \\
                              & SPR                                           & 0.210           & 0.218          & 0.194           & 0.471                   & 0.650          & 0.650          & 0.695          & 0.523          & \textbf{0.988} & \textbf{0.988} & \textbf{0.990} & 0.983          & 33.870 & 32.864 & 27.098 & 19.608                                       & 0.001 & 0.001  & 0.008 & 0.012              \\
                              & P2S                                           & 0.448           & 0.421          & 0.506           & 0.439                   & 0.526          & 0.551          & 0.539          & 0.535          & 0.968          & 0.968          & 0.957 & 0.967          & 4.719  & 4.117  & 2.011  & 4.535                                        & 0.093 & 0.115  & 0.098 & 0.125              \\
                              & DSE*                                          & 0.171           & 0.182          & 0.173           & 0.169                   & \textbf{0.710} & 0.674          & 0.691          & 0.697          & 0.987          & \textbf{0.988} & \textbf{0.990} & 0.988          & 0.956  & 2.267  & 2.254  & 0.873                                        & 0.339 & 0.207~ & 0.306 & 0.386              \\
                              & RVD/RPD                                       & ~\textbf{0.170} & 0.186          & \textbf{0.172}  & \textbf{0.146}          & 0.707          & \textbf{0.675} & \textbf{0.704} & \textbf{0.738} & 0.987          & 0.987          & 0.989          & \textbf{0.990} & 0.833  & 2.322  & 0.884  & \textbf{0.076}                               & 0.340 & 0.166  & 0.339 & \textbf{0.597}     \\ 
\midrule
\multirow{6}{*}{0.25\%-noise} & GD                                            & 1.173           & 1.092          & 1.796           & 1.152                   & 0.288          & 0.453          & 0.290          & 0.520          & 0.785          & 0.973          & 0.802          & 0.969          & 2.546  & 9.866  & 2.500  & 0.953                                        & 0.042 & 0.056  & 0.054 & 0.329              \\
                              & RIMLS                                         & 0.660           & 0.618          & 0.694           & 0.622                   & 0.339          & 0.355          & 0.316          & 0.259          & \textbf{0.970} & 0.971          & 0.957          & 0.943          & 8.510  & 3.746  & 1.703  & 2.549                                        & 0.044 & 0.055  & 0.002 & 0.053              \\
                              & SPR                                           & 0.383           & 0.326          & \textbf{0.340}  & 0.359                   & \textbf{0.524} & \textbf{0.531} & \textbf{0.484} & 0.412          & 0.968          & \textbf{0.975} & \textbf{0.970} & 0.960          & 15.918 & 31.800 & 12.833 & 31.422                                       & 0.009 & 0.003  & 0.009 & 0.003              \\
                              & P2S                                           & \textbf{0.359}  & 0.358          & 0.537           & 0.672                   & 0.498          & 0.482          & 0.432          & 0.389          & 0.964          & 0.964          & 0.929          & 0.934          & 3.637  & 6.649  & 2.417  & 1.429                                        & 0.079 & 0.068  & 0.050 & 0.067              \\
                              & DSE*                                          & 0.575           & 0.325          & 0.589           & 0.441                   & 0.302          & 0.432          & 0.299          & 0.530          & 0.795          &\textbf{ 0.975  }        & 0.792          & 0.968          & 2.562  & 2.534  & 2.436  & 2.367                                        & 0.038 & 0.066  & 0.042 & 0.075              \\
                              & RVD/RPD                                       & 0.484           & \textbf{0.316} & 0.456           & \textbf{0.\textbf{277}} & 0.355          & 0.505          & 0.370          & \textbf{0.559} & 0.818          & 0.973          & 0.831          & \textbf{0.976} & 2.551  & 10.448 & 2.512  & \textbf{0.190}                               & 0.051 & 0.068  & 0.053 & \textbf{0.430}     \\ 
\hline
\multirow{6}{*}{0.5\%-noise}  & GD                                            & 2.227           & 0.749          & 2.199           & 0.710                   & 0.154          & 0.383          & 0.155          & 0.312          & 0.664          & 0.975          & 0.682          & 0.963          & 2.618  & 21.794 & 2.554  & 0.977                                        & 0.020 & 0.015  & 0.025 & 0.179              \\
                              & RIMLS                                         & 0.645           & 0.687          & 1.732           & 0.653                   & 0.312          & 0.302          & 0.178          & 0.305          & \textbf{0.961} & 0.971          & \textbf{0.938} & 0.949          & 21.584 & 14.312 & 1.468  & 2.348                                        & 0.011 & 0.016  & 0.058 & 0.037              \\
                              & SPR                                           & 0.433           & 0.401          & \textbf{0.599}  & 0.382                   & 0.383          & \textbf{0.420} & \textbf{0.321} & 0.334          & 0.943          & \textbf{0.977} & 0.937          & 0.959          & 8.130  & 32.825 & 2.988  & 31.367                                       & 0.018 & 0.001  & 0.036 & 0.001              \\
                              & P2S                                           & \textbf{0.402}  & 0.428          & 0.873           & 0.451                   & \textbf{0.398} & 0.382          & 0.282          & 0.343          & 0.957          & 0.962          & 0.879          & 0.939          & 3.798  & 6.819  & 1.832  & 1.225                                        & 0.048 & 0.036  & 0.032 & 0.038              \\
                              & DSE*                                          & 1.663           & 0.464          & 1.674           & 0.371                   & 0.168          & 0.349          & 0.167          & 0.368          & 0.695          & 0.974          & 0.648          & 0.948          & 2.609  & 2.543  & 2.553  & 2.519                                        & 0.021 & 0.052  & 0.023 & 0.025              \\
                              & RVD/RPD                                       & 0.935           & \textbf{0.378} & 0.913           & \textbf{0.352}          & 0.255          & 0.409          & 0.262          & \textbf{0.446} & 0.732          & 0.976          & 0.741          & \textbf{0.968} & 2.573  & 16.881 & 2.537  & \textbf{0.215}                               & 0.034 & 0.029  & 0.038 & \textbf{0.301}     \\ 
\midrule
\multirow{6}{*}{1\%-noise}    & GD                                            & 6.114           & 1.031          & 5.891           & 2.838                   & 0.084          & 0.222          & 0.086          & 0.133          & 0.592          & 0.943          & 0.608          & 0.924          & 2.744  & 23.774 & 2.587  & 1.740                                        & 0.010 & 0.005  & 0.013 & 0.055              \\
                              & RIMLS                                         & 1.422           & 1.479          & 5.602           & 3.205                   & 0.207          & 0.197          & 0.090          & 0.097          & \textbf{0.957} & 0.945          & \textbf{0.903} & 0.906          & 27.173 & 6.731  & 1.532  & 3.454                                        & 0.002 & 0.007  & 0.024 & 0.009              \\
                              & SPR                                           & 1.014           & \textbf{0.816} & 2.126           & 2.420                   & 0.209          & 0.254          & 0.148 & 0.081          & 0.885          & \textbf{0.964} & 0.865          & 0.917          & 3.948  & 30.938 & 1.858  & 29.659                                       & 0.011 & 0.001  & 0.028 & 0.011              \\
                              & P2S                                           & \textbf{0.790}  & 0.779          & 2.934           & 3.051                   & \textbf{0.219} & \textbf{0.267} & 0.125          & 0.189          & 0.933          & 0.949          & 0.821          & 0.905          & 3.622  & 7.868  & 1.325  & 1.089                                        & 0.017 & 0.014  & 0.015 & 0.032              \\
                              & DSE*                                          & 5.401           & 1.471          & 5.412           & 2.129                   & 0.091          & 0.216          & 0.091          & 0.174          & 0.564          & 0.940          & 0.566          & 0.909          & 2.738  & 2.509  & 2.617  & 2.579                                        & 0.011 & 0.028  & 0.013 & 0.014              \\
                              & RVD/RPD                                       & 1.755           & 0.841          & \textbf{1.632 } & \textbf{1.564}          & 0.173          & 0.242          & \textbf{0.181}          & \textbf{0.241} & 0.669          & 0.949          & 0.669          & \textbf{0.926} & 2.584  & 12.705 & 2.515  & \textbf{1.045}                               & 0.022 & 0.013  & 0.026 & \textbf{0.119}     \\
\bottomrule
\end{tabular}
}
\end{table*}
\begin{figure*}[!htp]
	\centering
\vspace{-1mm}
\begin{overpic}
[width=.99\textwidth]{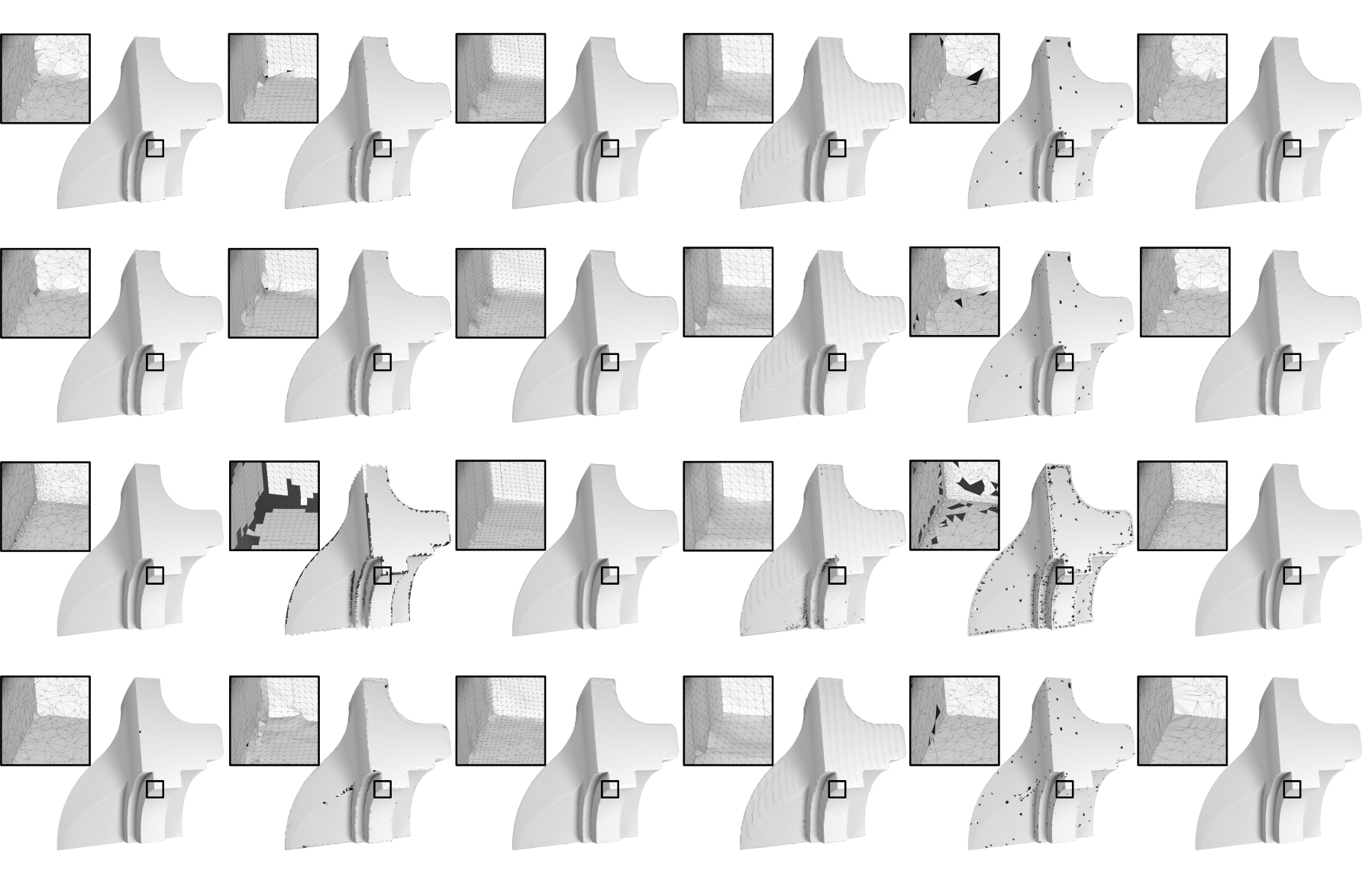}
\put(7,0.5){\textbf{Greedy}}
\put(24,0.5){\textbf{RIMLS}}
\put(41.5,0.5){\textbf{SPR}}
\put(58.5,0.5){\textbf{P2S}}
\put(74.5,0.5){\textbf{DSE}}
\put(87,0.5){\textbf{RVD/RPD(ours)}}
\put(-0.5,52){\rotatebox{90}{\textbf{w.o.}}}
\put(-0.5,34){\rotatebox{90}{\textbf{RIMLS}}}
\put(-0.5,20){\rotatebox{90}{\textbf{EAR}}}
\put(-0.5,4){\rotatebox{90}{\textbf{Ours}}}
\end{overpic}
\vspace{-3mm}
\caption{
Comparison with state of the arts of surface reconstruction from different point cloud consolidation methods. The whole pipeline of RFEPS surpasses other methods in terms of reconstruction fidelity and manifoldness.
}

	\label{FIG:Reconstruction}
\end{figure*}
\begin{figure*}[!htp]
	\centering
\vspace{-3mm}
\begin{overpic}
[width=.99\textwidth]{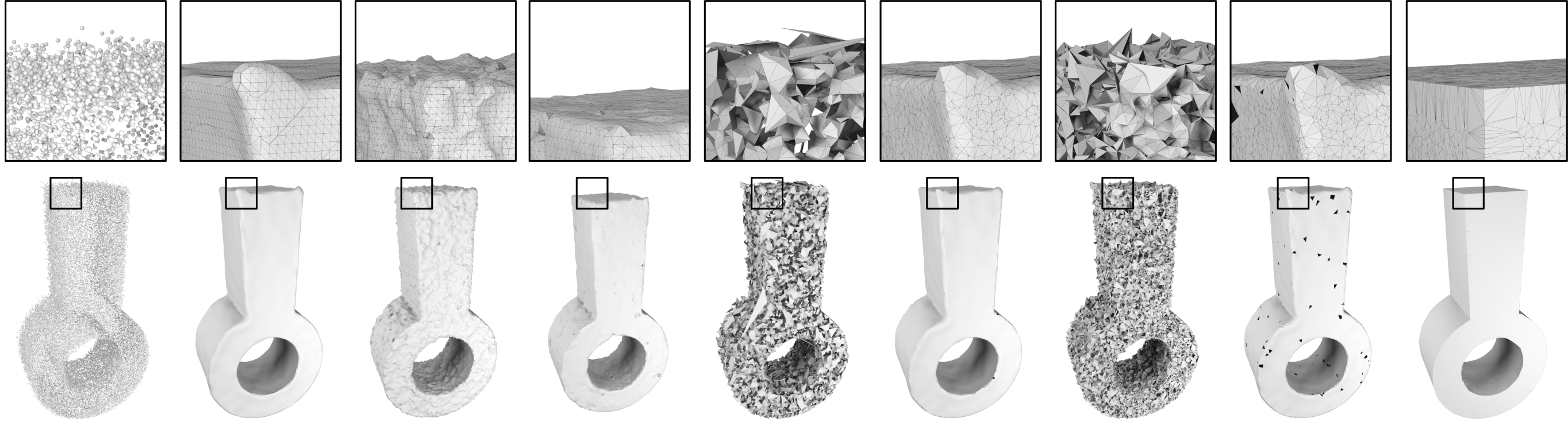}
\put(3.5,-1){\textbf{Input}}
\put(14,-1){\textbf{RIMLS}}
\put(26.5,-1){\textbf{SPR}}
\put(38,-1){\textbf{P2S}}
\put(49,-1){\textbf{GD}}
\put(57,-1){\textbf{RIMLS+GD}}
\put(71,-1){\textbf{DSE}}
\put(79,-1){\textbf{RIMLS+DSE}}
\put(93,-1){\textbf{Ours}}
\end{overpic}
\vspace{-1mm}
\caption{Comparison with state of the arts on a noisy point cloud input. RFEPS surpasses other methods in both the accuracy and manifoldness of the reconstruction.
}
\label{FIG:compareNoise}
\end{figure*}
\begin{figure*}[!htp]
	\centering
\vspace{-3mm}
\begin{overpic}
[width=\linewidth]{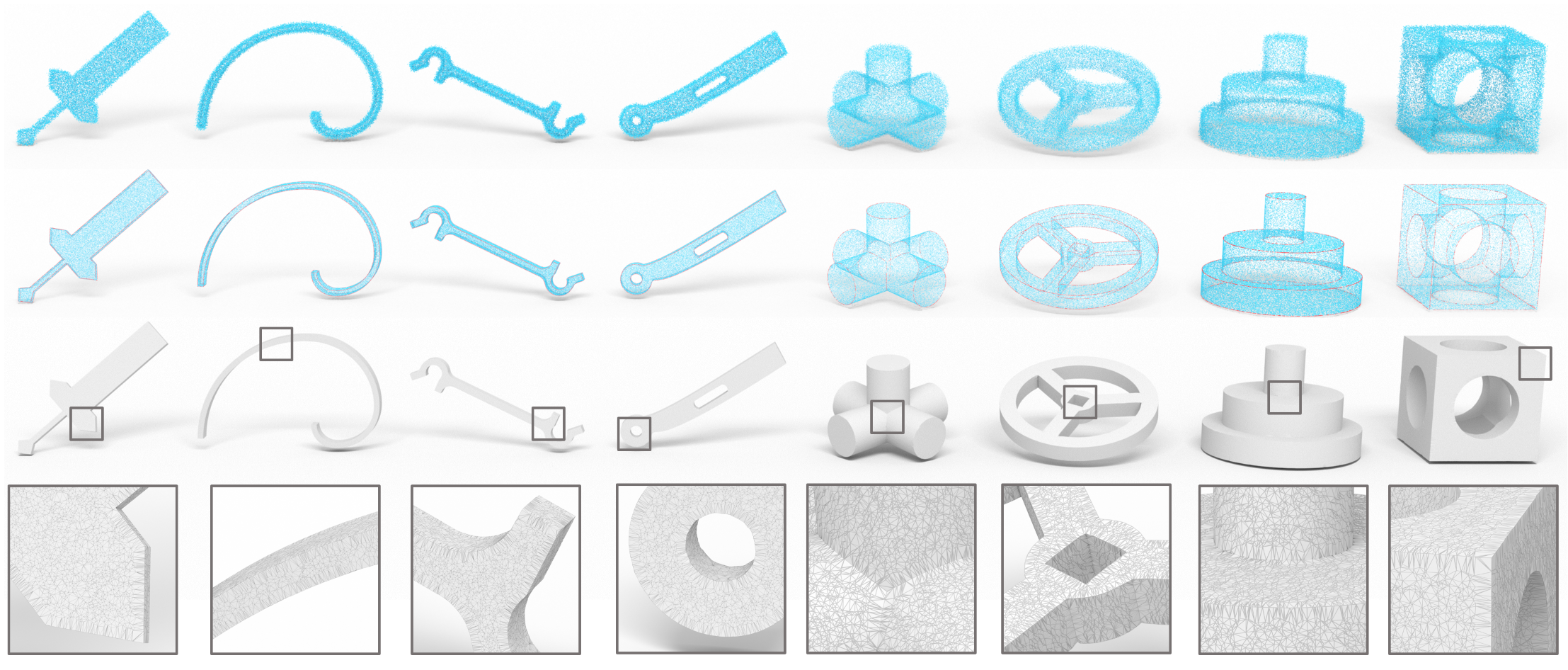}
\end{overpic}
\vspace{-8mm}
\caption{More reconstruction results produced by our algorithm pipeline.}
\vspace{-2mm}
\label{FIG:moreResults}
\end{figure*}
\subsection{Surface Reconstruction Quality}
The theme of this paper is 
to recover the underlying geometry, as well as the meshing, from a CAD-type point cloud. 
It includes a step of point consolidation and a step of mesh generation. 
Therefore, 
it is necessary 
to find the best combination of the point consolidation approach and surface reconstruction approach.
Table~\ref{tab:quantitative_noisefree}
gives the scores to evaluate various ``point consolidation plus surface reconstruction'' combinations 
on point data with various levels of noise,
where ``w.o.'' means ``without any point consolidation''.

The surface reconstruction solvers used for comparison include Greedy Delaunay (GD)~\cite{cohen2004greedy}, robust implicit moving least-square (RIMLS)~\cite{RIMLS}, Screened Poisson Reconstruction (SPR)~\cite{SPR}, Points2Surf (P2S)~\cite{P2S}, and DSE-meshing (DSE)~\cite{DSE}, where P2S, SPR, and RIMLS are implicit methods
while GD and DSE are interpolation-based. 
Note that the reconstruction strategy of the  restricted power diagram (RPD) proposed in this paper 
needs a base surface as the support.
We use the output of SPR to provide the base surface, but a different approach could also work. 
In addition, all the newly added points by our point consolidation strategy have an edge-point label, which enables us to set a larger weight in computing the RPD (see Section~\ref{subsec:RPD_recon}). The consolidated point clouds by RIMLS and EAR lack the edge-point label, and thus we have to use the RVD instead (all the weights are equal to each other). 
Our reconstruction pipeline includes a step of regularizing normal vectors (\SQ{see Eq.~(\ref{eq:denoise_init}) and Eq.~(\ref{eq:three})}) and thus does not require the input point cloud to be equipped with reliable normal vectors.
It's worth noting that 
the given point cloud may lack normal information. 
Under the circumstance,
for those approaches that require normal information,
we use PCA to initialize the normal vectors.

\begin{table}[!t]
\caption{Running time (in seconds) w.r.t the number of points \#V. We use the block model shown in Figure~\ref{fig:teaser} for test.
}
\label{table:time}
\vspace{-4mm}
\resizebox{.87\linewidth}{!}{
\begin{tabular}{p{0.7cm}|p{1.2cm}<{\centering}|p{1.2cm}<{\centering}|p{1.2cm}<{\centering}|p{1.2cm}<{\centering}|p{1.2cm}<{\centering}}
\toprule
\#V   & 10K    & 30K     & 50K     & 70K     & 100K        \\ \midrule
T1     & 1.13  & 0.94   & 2.17  & 3.32   & 5.15 \\
T2    & 0.39  & 0.93  & 1.94 & 2.70 & 3.55 \\
T3    & 0.05  & 0.14   & 0.21  & 0.28   & 0.37  \\
T4    & 0.16  & 0.34  & 0.81 & 1.21 & 1.76  \\
T5    & 0.02  & 0.05   & 0.10  & 0.13   & 0.17  \\  \midrule
SPR   & 1.38  & 1.73   & 2.89  & 3.68   & 4.08  \\
RPD    & 0.50  & 1.01   & 1.41  & 1.85   & 2.59  \\ \midrule
Total & 3.63 & 5.14 & 9.53 & 13.17 & 17.67 \\
\bottomrule
\end{tabular}
}
\label{tab:time}
\vspace{-6.5mm}
\end{table}

\begin{figure*}
	\centering
\begin{overpic}
[width=\linewidth]{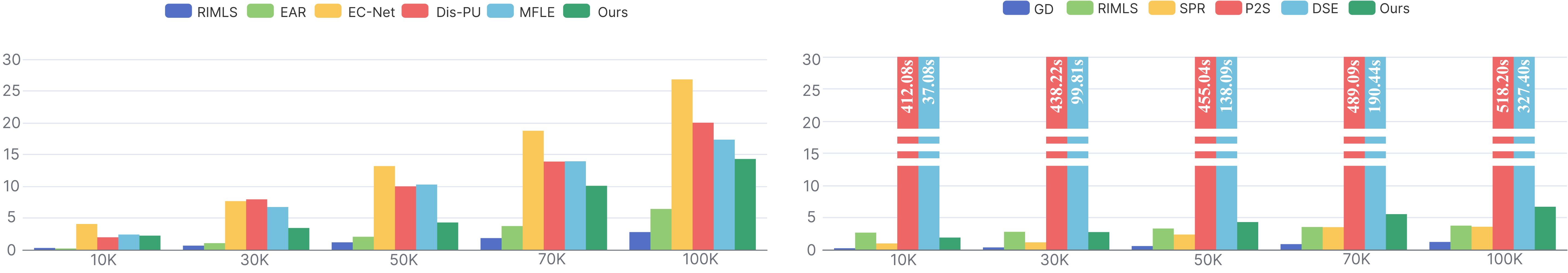}
\put(15,-1.5){(a)~Point Cloud Consolidation}
\put(67,-1.5){(b)~Surface Reconstruction}
\end{overpic}
\vspace{-4mm}
\caption{Running time (in seconds) of point cloud consolidation and surface mesh reconstruction. Our method is better than or comparable to existing methods.}
	\label{FIG:timecompare}
\vspace{-4mm}
\end{figure*}
The statistics in Table~\ref{tab:quantitative_noisefree}
enable us to make some observations. 
First, our reconstruction pipeline, 
including the edge-point augmentation strategy proposed in Section~\ref{sec:init}-\ref{sec:FeaturePoint}
and the RPD based reconstruction  strategy proposed in Section~\ref{subsec:RPD_recon},
is the best combination - its ECD and EF1 scores are significantly better than other combinations, and its CD/F1/NC scores are better than or comparable to other combinations. 
Second, it is hard for implicit methods to retain feature lines in the output mesh (see the ECD and EF1 scores in Table~\ref{tab:quantitative_noisefree}). 
For example, RIMLS cannot produce sharp feature lines.
Figure~\ref{FIG:Reconstruction} visualizes 
the reconstruction results for all the
``point consolidation plus surface reconstruction'' combinations.
But we also point out that
the existing interpolation-based approaches, such as GD and DSE,
\SQ{cannot rigorously guarantee a watertight manifold output,
especially when the given point cloud contains thin structures.} 
DSE cannot even guarantee the face-orientation consistency, so we have to use additional post-processing~\cite{takayama2014simple}.
\SQ{In contrast, our RPD reconstruction inherits the nice features of the RVD 
and has advantages in guaranteeing manifoldness.}
Figure~\ref{FIG:compareNoise} shows
the results of some typical reconstruction approaches on a noisy point cloud, 
either directly running on the original data (e.g., SPR)
or calling a compositional solution (e.g., RIMLS+GD, RIMLS+DSE). 
More reconstruction results of CAD models by our algorithm pipeline are shown in Figure~\ref{FIG:moreResults}.
We provide more detailed visual comparison results in Appendix~A.

\subsection{Runtime Performance}
\label{sec:time_efficiency}
We report the statistics about running time in Table~\ref{tab:time}, where the number of points \#V ranges from 10K to 100K. 
Our algorithm consists of multiple stages,
and we record the running time for each step:
T1: point cloud denoising; 
T2: edge zone identification;
T3: normal-vector regularization;
T4: point-location refinement;
T5: edge-point generation;
SPR: using the SPR to construct the base surface;
RPD: using the RPD to generate the final triangle mesh. 
Considering the steps of edge zone identification,
normal-vector regularization and edge-point generation can be parallelized, we use 24 threads to speed up the computation of the three steps. 
It can be seen from the statistics that 
the running time of each stage increases  linearly w.r.t.~\#V
and T1 is the most time-consuming stage.
In our experiments, 50K is the default size of the input point cloud, which requires about $10$ seconds. Detailed statistics comparing the run-time performance of different approaches are shown in Figure~\ref{FIG:timecompare}.
It can be seen that our algorithm has a competitive run-time performance, especially compared with deep learning techniques.


\begin{figure}[!t]
	\centering
\begin{overpic}
[width=.9\linewidth]{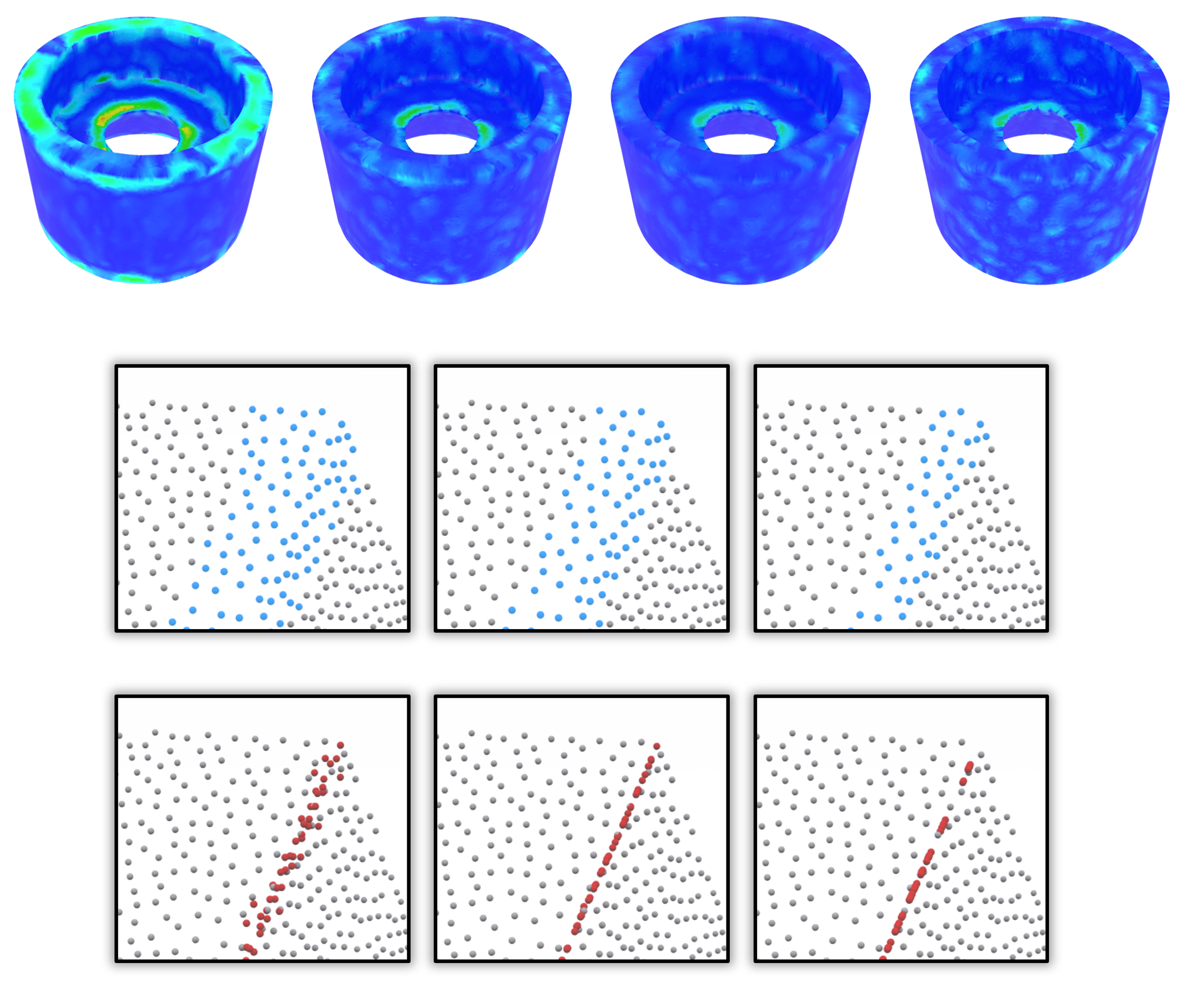}
\put(82,57){$\xi=1.0$}
\put(57.5,57){$\xi=0.1$}
\put(31,57){$\xi=0.01$}
\put(5,57){$\xi=0.001$}
\put(16,0){$\mu=1.0$}
\put(42,0){$\mu=0.01$}
\put(68,0){$\mu=0.001$}
\put(17,28){$r=3\delta$}
\put(44,28){$r=2\delta$}
\put(70,28){$r=1.5\delta$}
\end{overpic}
\vspace{-2mm}
\caption{Influence of different parameter settings. 
Top row: $\xi$ is used to tune the degree of denoising. 
Middle row: the patch radius~$r$ influences the width of the feature-line zone.
Bottom row: $\mu$ is used to avoid the drift too far away from its original position. }
	\label{FIG:Parameter}
\vspace{-3mm}
\end{figure}
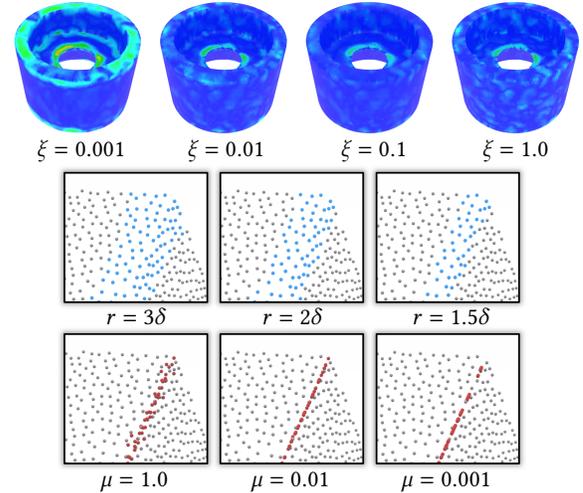

\begin{table}[!t]
\centering
\caption{Influence of the parameters $\xi$, $r$ and $\mu$ on the reconstruction quality. We report the error relative to the default configurations.
}
\label{tab:ablation}
\vspace{-3.5mm}
\resizebox{0.40\textwidth}{!}
{
\begin{tabular}{p{1.2cm}|p{1.5cm}<{\centering}|p{0.7cm}<{\centering}|p{0.75cm}<{\centering}|p{1.7cm}<{\centering}|p{0.75cm}<{\centering}}
\toprule
 & $\mathrm{CD}\left(\times 10^{5}\right) \downarrow$ & $\mathrm{F1} \uparrow$ & $\mathrm{NC} \uparrow$ & $\mathrm{ECD}\left(\times 10^{2}\right) \downarrow$ & $\mathrm{EF1} \uparrow$ \\ \midrule
$\xi={1.0}$ & 1.001 & 0.999 & 0.997 & 1.000 & 0.992 \\
$\xi={0.1}$ & 1.000 & 1.000 & 1.000 & 1.000 & 1.000 \\
$\xi={0.01}$ & 1.033 & 0.999 & 0.955 & 1.090 & 0.973\\
$\xi={0.001}$ & 1.106 & 0.998 & 0.997 & 1.093 & 0.835 \\
\midrule
$r={1.5\delta}$ & 1.010 & 1.001 & 0.995 & 1.020 & 0.981 \\
$r={2\delta}$ & 1.000 & 1.000 & 1.000 & 1.000 & 1.000 \\
$r={3\delta}$ & 0.997 & 1.001 & 0.996 & 1.038 & 0.970 \\
\midrule
$\mu={0.001}$ & 1.005 & 0.999 & 1.001 & 1.035 & 0.974 \\
$\mu={0.01}$ & 1.000 & 1.000 & 1.000 & 1.000 & 1.000 \\
$\mu={1.0}$ & 1.021 & 0.999 & 0.997 & 1.043 & 0.925 \\ \bottomrule
\end{tabular}
}
\vspace{-4mm}
\end{table}

\begin{figure}[!t]
\vspace{-5mm}
	\centering
\begin{overpic}
[width=.9\linewidth]{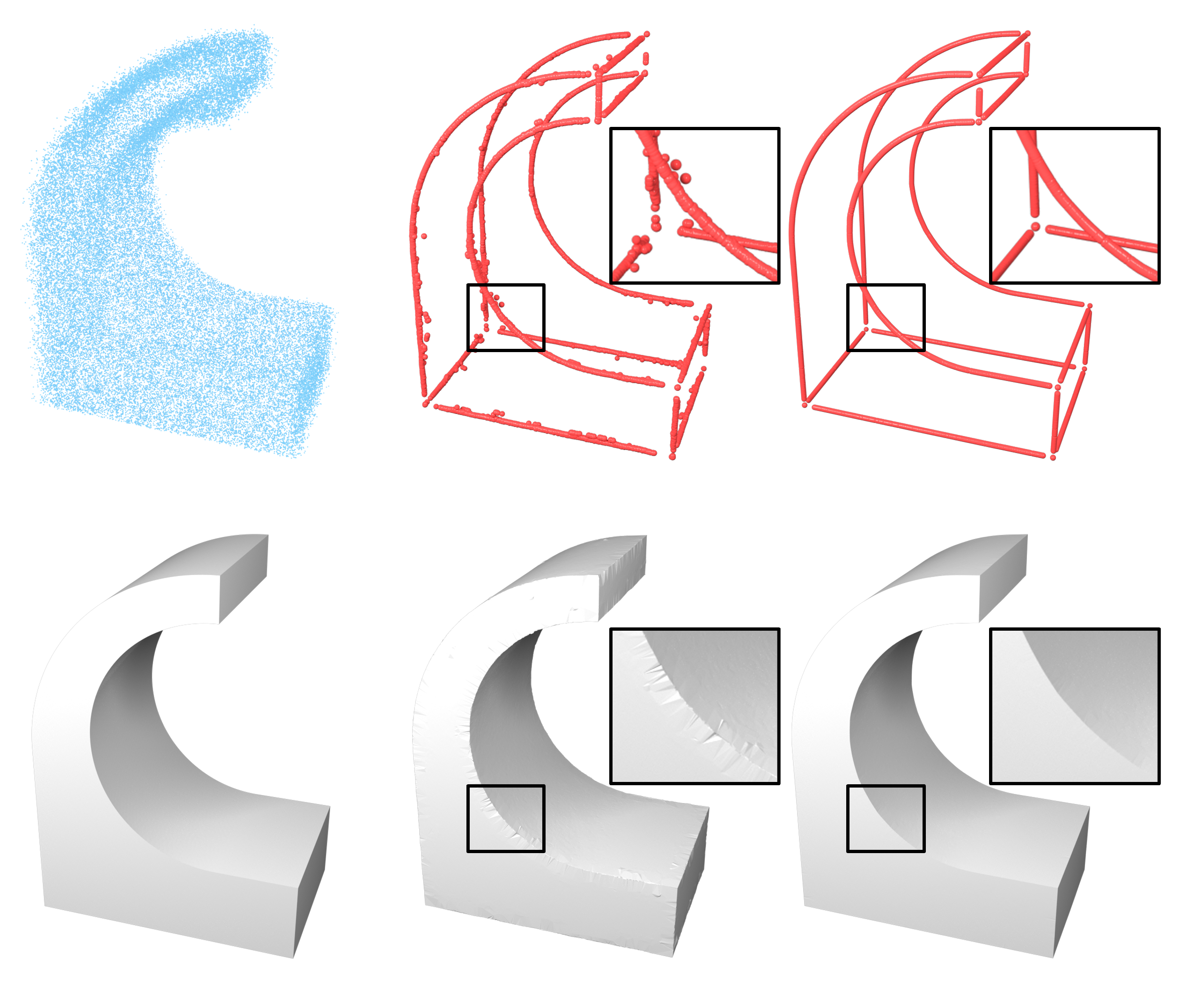}
\put(7,42.5){(a) Input}
\put(36,42.5){(b) K-means}
\put(38,-4){(EF1: 0.275)}
\put(72,42.5){(c) Ours}
\put(72,-4){(EF1: 0.551)}
\put(9,0){(d) GT}
\put(33,0){(e) K-means Recon}
\put(67,0){(f) Ours Recon}
\end{overpic}
 \vspace{-1mm}
\caption{
\SQ{Visual comparison between k-means~\cite{hartigan1979algorithm} and our formulation (Eq.~(\ref{eq:edge_zone}) and Eq.~(\ref{eq:three})).}
}
\label{FIG:kmeans}
\vspace{-4mm}
\end{figure}

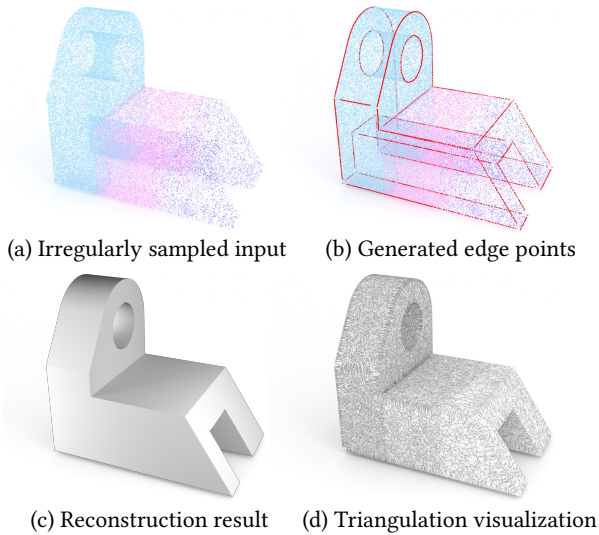
\begin{figure}[!t]
\vspace{-0.5mm}
	\centering
\begin{overpic}
[width=.9\linewidth]{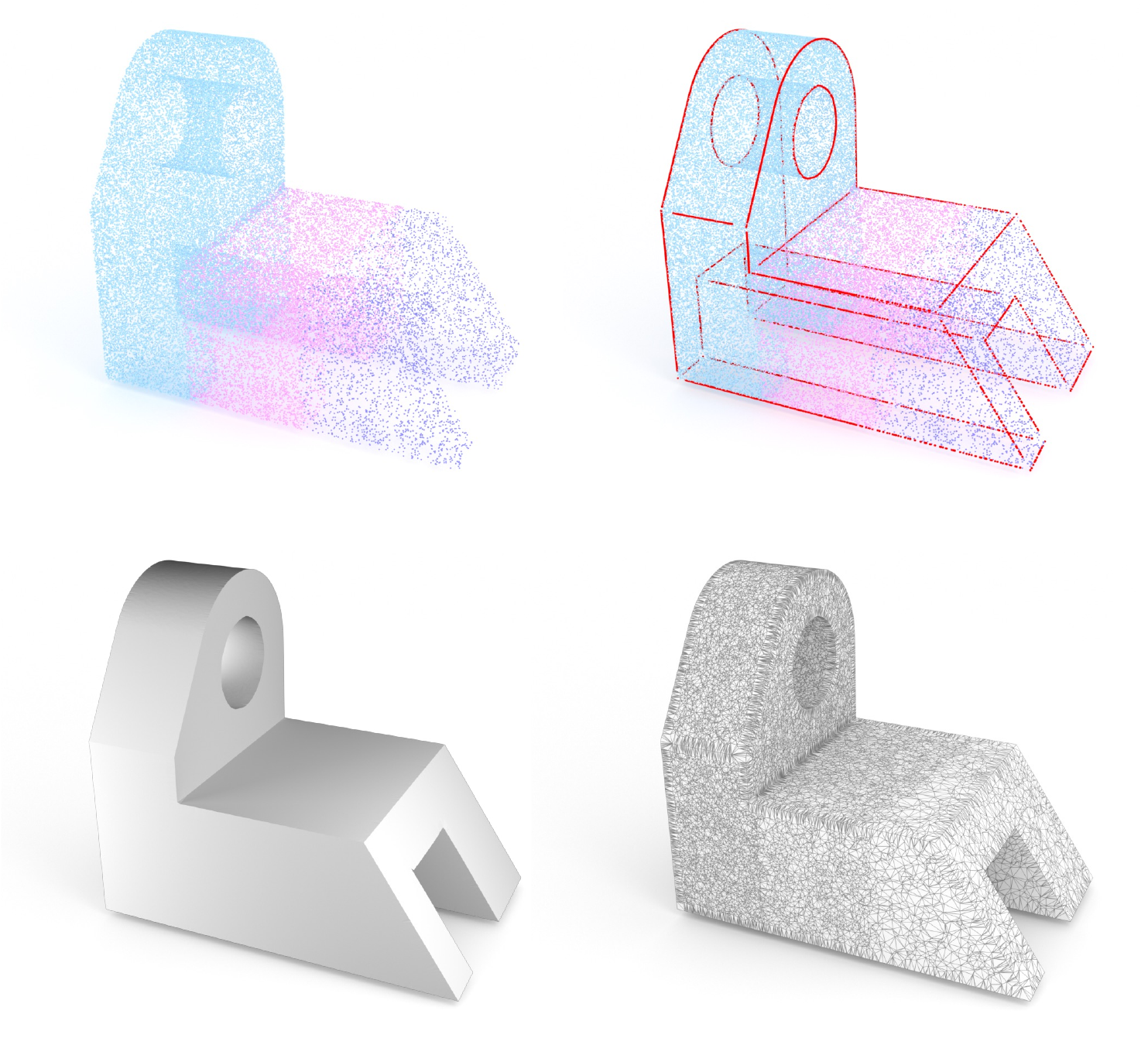}
\put(1,47){(a) Irregularly sampled input}
\put(56,47){(b) Generated edge points }
\put(5,0){(c) Reconstruction result}
\put(52,0){(d) Triangulation visualization}
\end{overpic}
 \vspace{-2mm}
\caption{
We prepare a synthetic point cloud
with varying point density (a),
where the density is visualized in a color-coded style.
Our algorithm works well 
for the irregularly sampled data
and can generate faithful edge points (b).
By running the RPD, our algorithm 
is able to yield a polygonal mesh that 
manifests the real geometry (c),
whose triangulation is visualized in (d).
}
\label{FIG:DiffSampling}
\vspace{-6mm}
\end{figure}

\subsection{Influence of Parameters}
\label{sec:parameters}
Figure~\ref{FIG:Parameter} visualizes the influence of the parameters $\xi$, $r$, and $\mu$.
The parameter $\xi$ in Eq.~(\ref{eq:denoise_init}) is used to control the denoising degree. 
A large $\xi$ tends to suppress the movement of a point 
whereas a small $\xi$ tends to encourage the smoothness of point locations and normal vectors. 
We take $\xi=0.1$ in our experiments.
Our algorithm pipeline is insensitive to $\xi$
since the pipeline includes further 
normal-vector regularization (Step 3)
and point-location refinement (Step 4). 
The patch radius $r$ is used to control the width of the edge zone.
If $r$ is too small, it is likely that too few points participate in the optimization (see Eq.~(\ref{equa:FeatureZone},\ref{equa:6})), leading to the misclassification of points. 
But if $r$ is too large, our algorithm may fail for thin-plate models. 
Figure~\ref{FIG:Parameter} shows three different results by setting $r=3\delta,2\delta,1.5\delta$, where $\delta$ is the average gap between points. In our experiments, we set $r=2\delta$ by default.
In the step of edge-point generation,
we project a point in the edge zone onto the nearby geometry edge.
The optimization of Eq.~(\ref{equa:6}) includes a term $\mu \|z_i-p_i\|^2$ to avoid point drifting along the potential feature line. 
We take $\mu = 0.01$ by default. 
Table~\ref{tab:ablation}
gives the statistics about how different parameter settings influence the reconstruction quality.

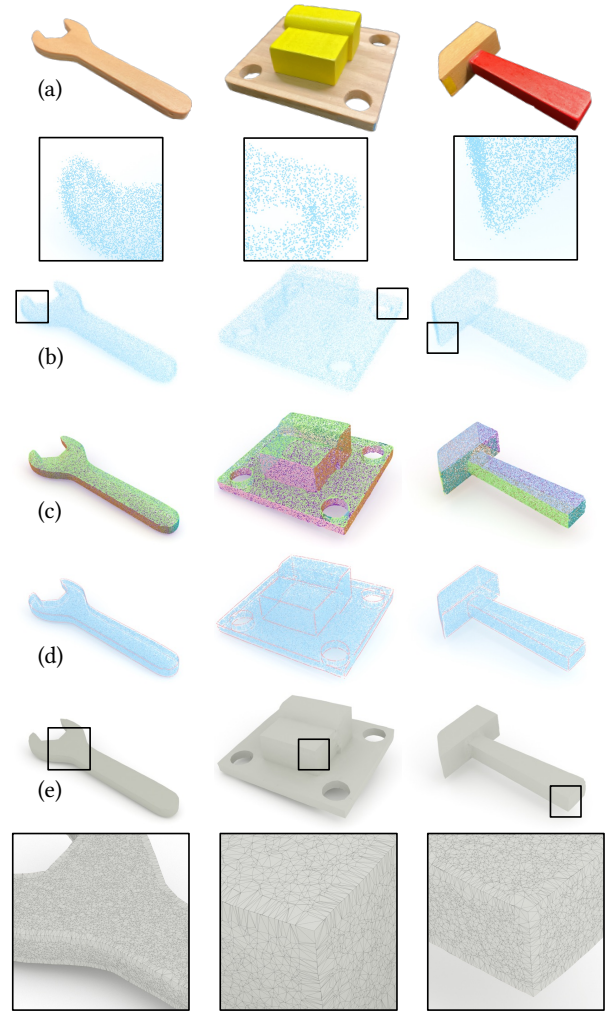
\begin{figure}[t]
	\centering
\vspace{-5mm}
\begin{overpic}
[width=\linewidth]{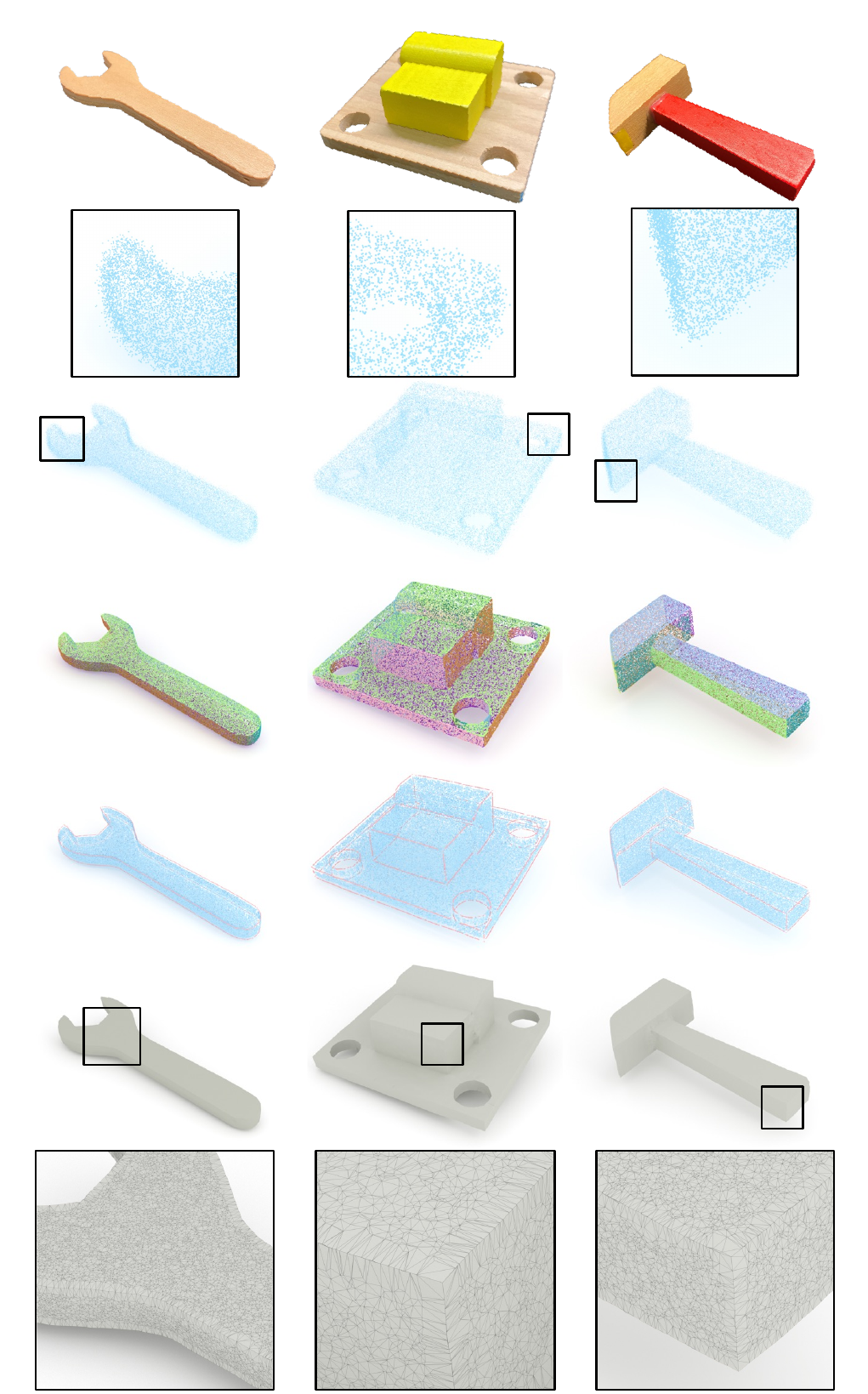}
\put(5,89){(a)}
\put(5,63){(b)}
\put(5,48){(c)}
\put(5,34){(d)}
\put(5,21){(e)}
\end{overpic}
\vspace{-4mm}
\caption{Test RFEPS on real-life scans. (a) Real-world shapes. (b) Raw scans. (c) Recovered normal vectors. (d) Predicted edge points. (e) Reconstructed surfaces.} 	
\vspace{-6mm}
\label{FIG:real_scan}
\end{figure}

\subsection{Optimal Mass Transport v.s. K-means}
\label{sec:kmeans_init}
\SQ{
In this paper, we formulate the task of edge-zone identification
as optimal mass transport; See Eq.~(\ref{eq:edge_zone}).
We further use Eq.~(\ref{eq:three}) to regularize normal vectors. 
Both of them are implemented based on optimization. 
Although k-means seemingly works, 
the contrast visualized in Figure~\ref{FIG:kmeans}
shows that our approach can generate feature lines with higher fidelity,
due to the ability of our optimization driven formulation
to more accurately characterize the geometric properties of a geometry edge. 
It's worth noting that 
$k=2$ for substituting k-means for Eq.~(\ref{eq:edge_zone})
and $k=3$ for substituting k-means for Eq.~(\ref{eq:three}).
}

\subsection{Robustness to Point Density}

To the best of our knowledge, 
most of the existing reconstruction approaches are sensitive
to point density. 
To test the robustness to point density, we prepare a synthetic point cloud
with varying point density as shown in Figure~\ref{FIG:DiffSampling}(a).
It can be seen that
our reconstruction algorithm pipeline
can deal with the irregularly sampled data
(see Figure~\ref{FIG:DiffSampling}(b-d)),
because
the choice of the parameters $\xi, r,\mu$
can adapt to the variation of point density. 
For example, the value of $r(=2\delta)$
is related to the average gap between points.

\label{ap:pc_density}

\begin{figure}[!htp]
	\centering
\vspace{-3mm}
\begin{overpic}
[width=.95\linewidth]{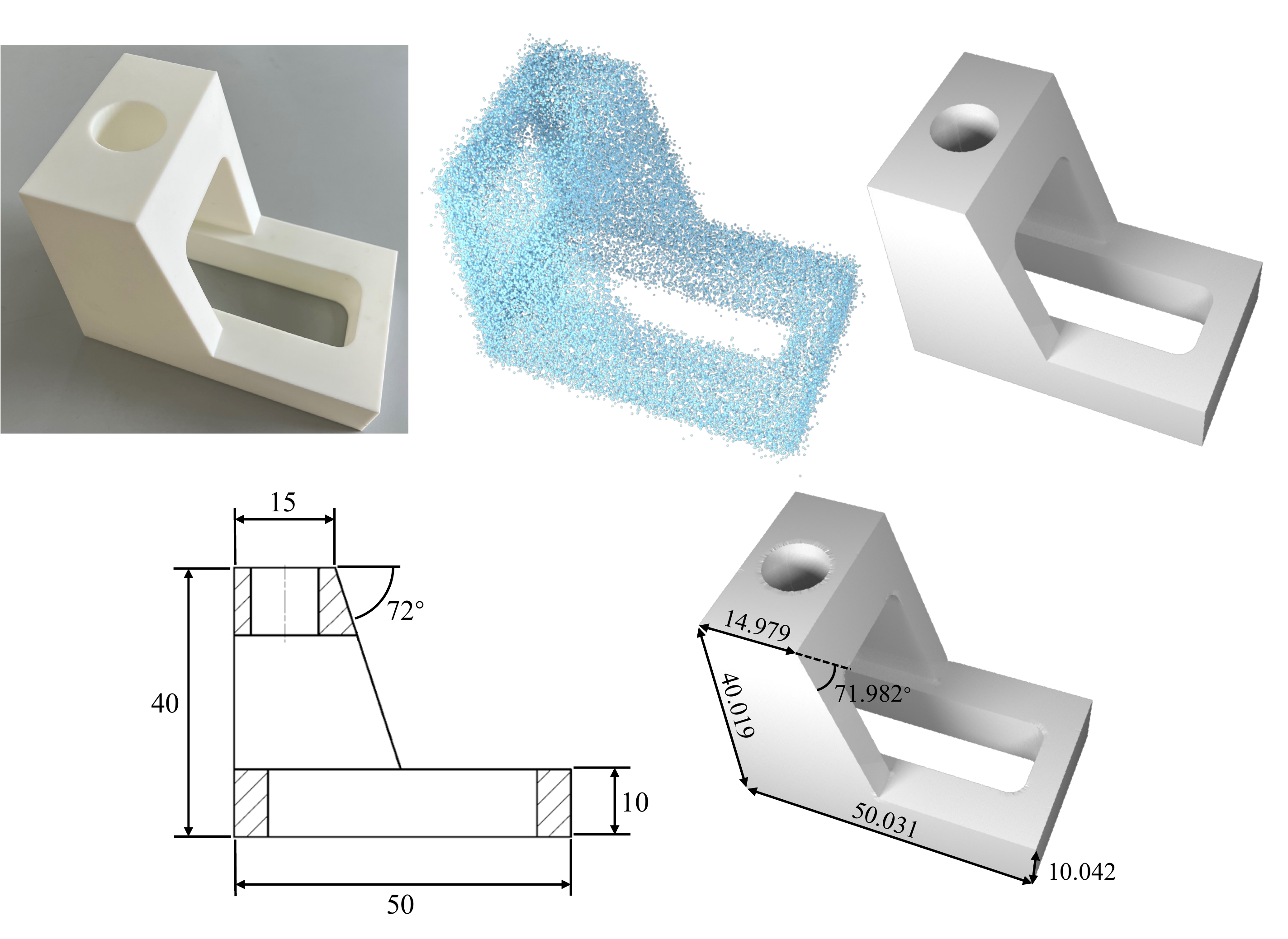}
\put(13,35){(a)}
\put(46,35){(b)}
\put(79,35){(c)}
\put(28.5,-1.5){(d)}
\put(66,-1.5){(e)}
\end{overpic}
\vspace{-2mm}
\caption{Given a raw scan of a workpiece,
	RFEPS is able to reproduce the initial design intent and the specifications of the original project. 
	The model is downloaded from~\cite{3dmodel}.}
	\vspace{-2mm}
\label{FIG:3dprint}
\end{figure}

\begin{figure}[t]
	\centering
\vspace{-3mm}
\begin{overpic}
[width=.95\linewidth]{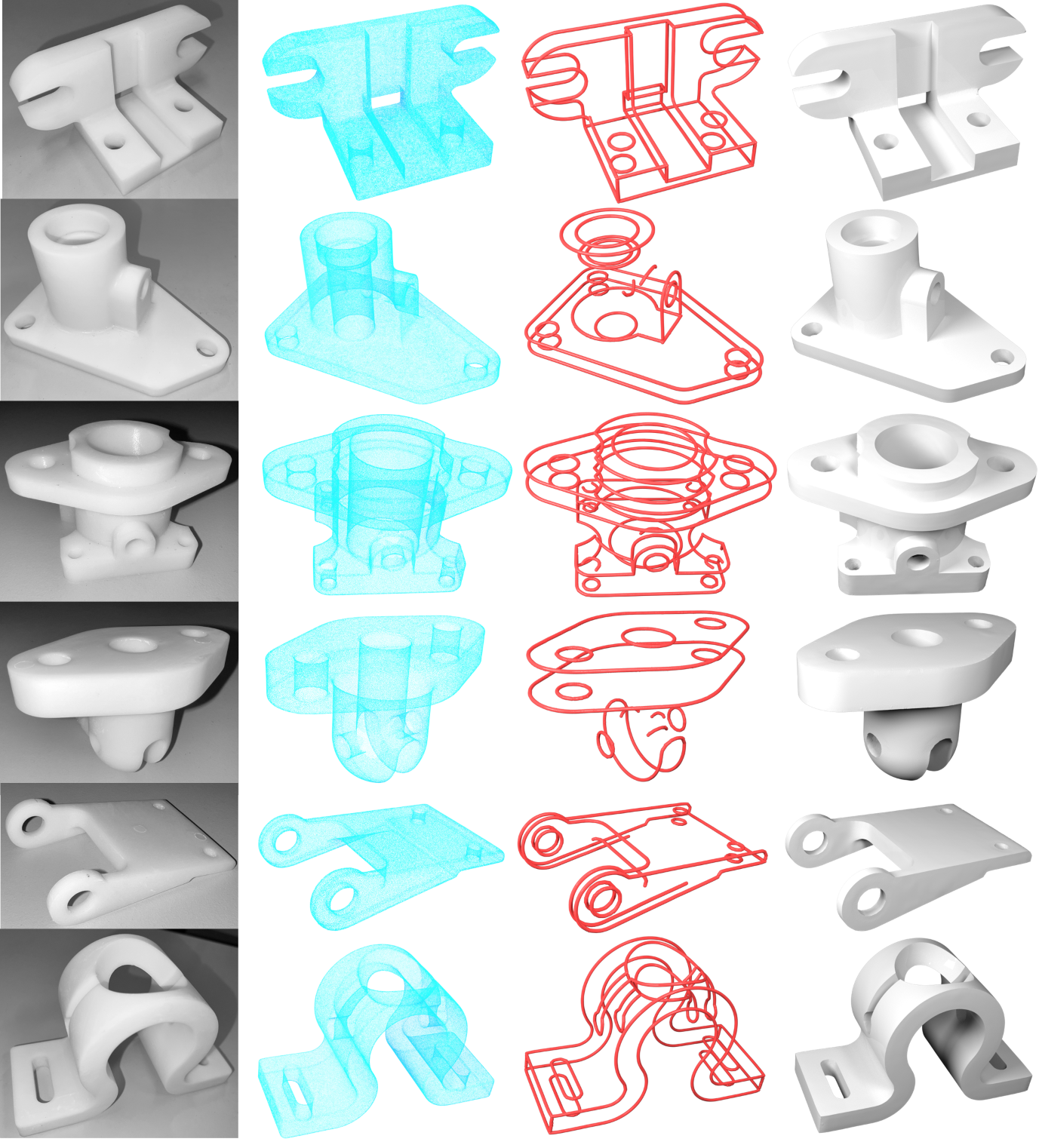}
\put(9,-2){(a)}
\put(32,-2){(b)}
\put(55.5,-2){(c)}
\put(79,-2){(d)}
\end{overpic}
\vspace{-1.5mm}
\caption{\SQ{More results from real scans. (a) Real-life objects. (b) Raw scans from SIMSCAN scanner~\cite{scanner2}. (c) Feature points produced by RFEPS. (d) Reconstructed feature-line equipped polygonal surface.}}
\vspace{-6mm}
\label{FIG:real_scan2}
\end{figure}
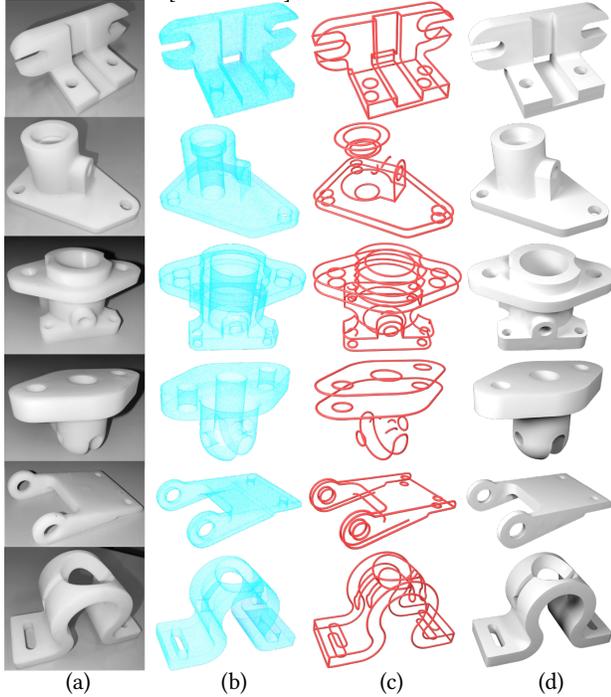

\subsection{Results on Real Scans}
\label{sec:real_scan}
\SQ{We scanned several real-life objects with a desktop scanner EinScan SE~\cite{scanner} ($0.1\text{mm}$ accuracy).} As Figure~\ref{FIG:real_scan} shows, the input points have an inhomogeneous distribution with conspicuous noise;
See the close-up views in the second row. 
Furthermore, quite few points are located on the edge of the geometry. 
Despite this, our algorithm can effectively eliminate noise, regularize normal vectors (see  Figure~\ref{FIG:real_scan}(c)),
and infer the precise locations of edge points (see Figure~\ref{FIG:real_scan}(d)).
Our algorithm is specially designed for CAD-type point data and each step of the algorithm fully considers 
the features of CAD models, particularly that the normal vectors have an abrupt change across the feature lines.  
The reconstructed results show that 
RFEPS can produce a high-fidelity reconstructed surface with neat feature lines, which validates the robustness and usefulness.

We further demonstrate the capability of our method in accurately reproducing the initial design intent and the specifications of the original project. 
For validation, 
we manufacture a model shown in Figure~\ref{FIG:3dprint} with a 3D printer, scan it into a point cloud, and then reconstruct it into a feature-line equipped polygonal surface by our RFEPS. 
It can be seen from Figure~\ref{FIG:3dprint}(d) and Figure~\ref{FIG:3dprint}(e) that the reconstruction 
dimensions are very close to the original design, 
which validates the usefulness in reconstructing a CAD model. 

\SQ{
Furthermore, we use a different scanner~\cite{scanner2} (SIMSCAN; $0.02\text{mm}$ accuracy)
to obtain raw scans of CAD models.
The number of points ranges from 100K to 600K.
As shown in Figure~\ref{FIG:real_scan2}, RFEPS not only accurately recovers the edge points but also achieves high quality reconstruction result, which validates the effectiveness of our method.
}

\subsection{Potential Applications}
\label{sec:potential_application}
\begin{figure}[!t]
	\centering
	\vspace{-2mm}
\begin{overpic}
[width=.85\linewidth]{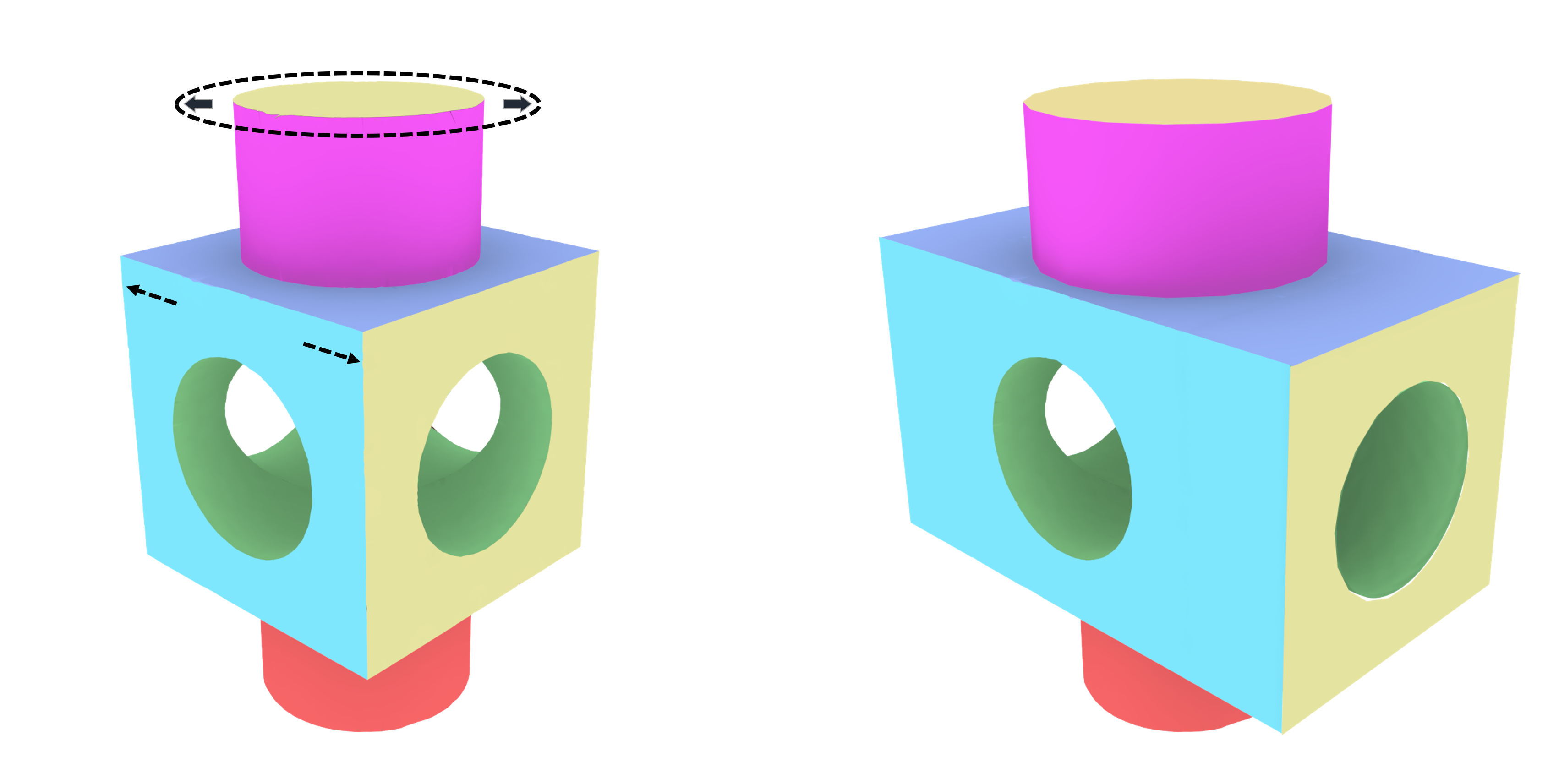}
\put(17.5,-1){(a)}
\put(73.5,-1){(b)}
\end{overpic}
	\vspace{-2mm}
\caption{\SQ{Flexible model editing with the support of feature lines. (a)~resizing instructions. (b)~the resized result.}}
	\label{FIG:CAD}
		\vspace{-2mm}
\end{figure}

The biggest benefit of recovering a feature-line equipped model lies in supporting various model edit tasks such as locally resizing a model. 
Upon the surface being reconstructed, it is easy to decompose the whole surface into a set of surface patches. As Figure~\ref{FIG:CAD} shows, each facet of the CAD surface is either planar or cylindrical.
\SQ{With the prior knowledge about surface types, the detailed implicit equation of each facet can be fitted following~\cite{du2021boundary}, which enables one to quickly estimate the driven parameters for defining each surface primitive.}
Therefore, users 
are allowed to specify different parameters to resize the shape locally.


%% file: Conclusion.tex
\begin{figure}[!t]
	\centering
 \vspace{-2mm}
\begin{overpic}
[width=\linewidth]{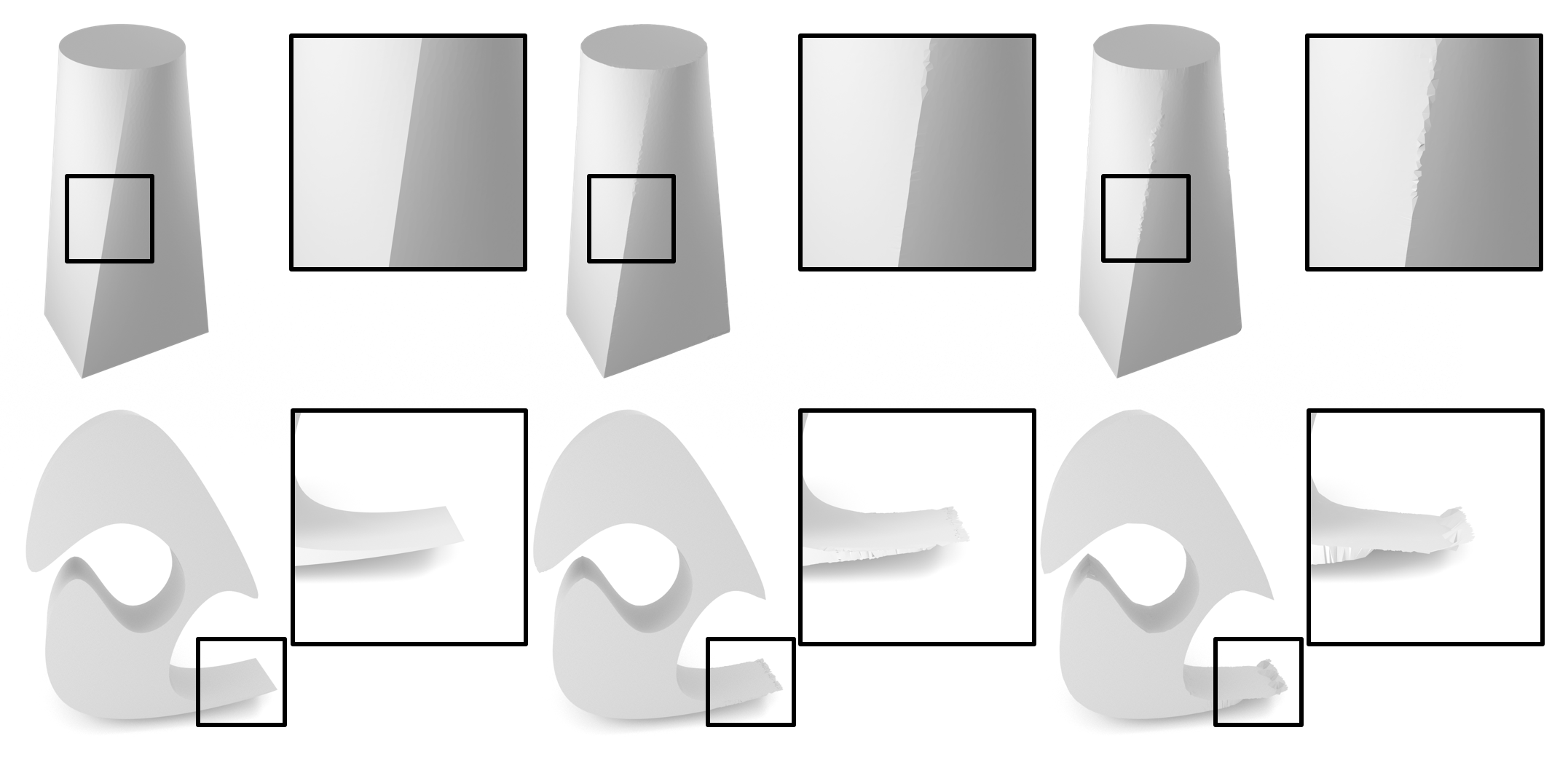}
\put(8,-1){\textbf{GT}}
\put(37,-1){\textbf{no-noise}}
\put(66.5,-1){\textbf{1.0\% noise}}
\end{overpic}
\vspace{-6mm}
\caption{\SQ{RFEPS may fail on a model with extremely small dihedral angles.}}
\label{FIG:FailureCase2}
\vspace{-4mm}
\end{figure}

\section{Limitation}
\label{sec:limitation}
\paragraph{Dihedral angle}
REFPS, in its current form, supports a dihedral-angle range of $[\pi/6, 5\pi/6]$. If the dihedral angle is too close to $\pi$ (approximately planar),
it is hard for Eq.~(\ref{equa:6}) to infer the precise projection position of a point in the edge zone due to numerical issues; See the top row of Figure~\ref{FIG:FailureCase2}.
If the dihedral angle is too close to $0$ (a sharp turn)
or the model contains a very thin part, 
the RPD may produce misconnections between two points on different sides;
See the bottom row of Figure~\ref{FIG:FailureCase2}.

\begin{figure}[!t]
	\centering
\begin{overpic}
[width=.9\linewidth]{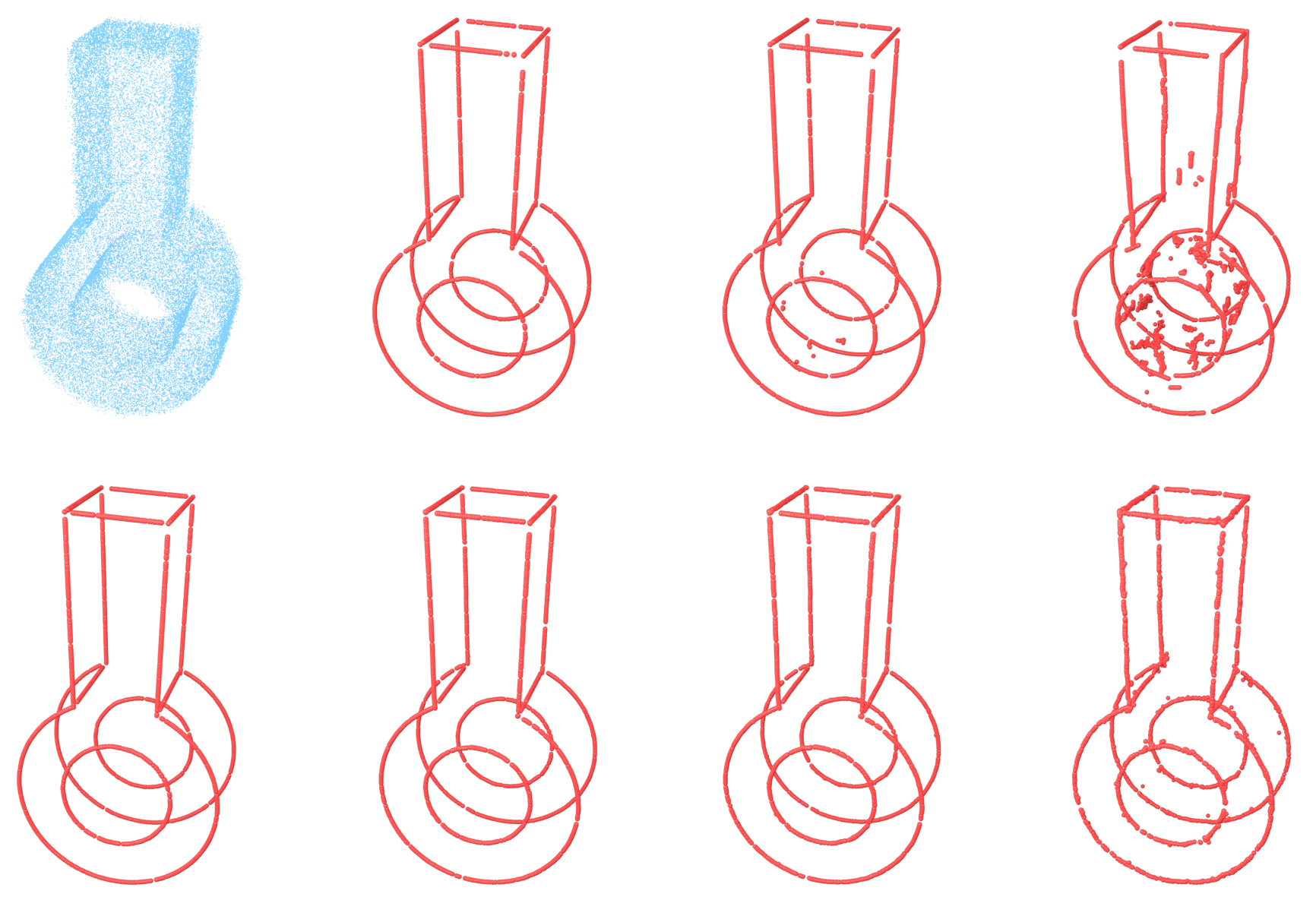}
\put(5.5,35){\textbf{Input}}
\put(29,35){\textbf{5\% reverse}}
\put(54.7,35){\textbf{10\% reverse}}
\put(81.4,35){\textbf{20\% reverse}}
\put(3.5,-1.5){\textbf{no-noise}}
\put(30,-1.5){\textbf{5\% noise}}
\put(55,-1.5){\textbf{10\% noise}}
\put(82.4,-1.5){\textbf{20\% noise}}
\end{overpic}
\vspace{-3mm}
\caption{\SQ{
Test the robustness to normal inconsistency and noise
by observing the quality of the augmented edge points.
Top row: We randomly reverse 5\%, 10\%, 20\% normal vectors, respectively. 
Bottom row: We add 5\%, 10\%, 20\% noise  to normal vectors, respectively.}}
\label{FIG:FailureCase_normal}
\vspace{-4mm}
\end{figure}

\paragraph{Normal inconsistency and noise}
\SQ{
Note that our multi-stage framework
has two steps of refining normal vectors. 
It is necessary to test if our algorithm 
highly depends on normal consistency and accuracy. 
In Figure~\ref{FIG:FailureCase_normal},
the top row shows the augmented edge points when one randomly flips 
5\%, 10\% and 20\% normal vectors,
while the bottom row shows the augmented edge points when one 
adds white-noise perturbance to normal vectors before edge-point generation (see Eq.~\ref{equa:6}):
$$
\mathbf{n} = \frac{\mathbf{n} + \tau \mathbf{n}_{\text{rand}}}{\|\mathbf{n} + \tau \mathbf{n}_{\text{rand}}\|},
$$
where $\mathbf{n}_{\text{rand}}$ is a random unit vector,
and $\tau = 5\%, 10\%, 20\%$ respectively. 
}
\XR{As a multi-stage framework, normal consistency influences Step 2 and Step 3 during point cloud consolidation.}
\SQ{However, as shown in Figure~\ref{FIG:FailureCase_normal}, RFEPS is insensitive to normal inconsistency.
It fails only if the point cloud has severe normal inconsistency, i.e., a 20\% reverse percentage.  }

\begin{figure}[!htp]
	\centering
\begin{overpic}
[width=\linewidth]{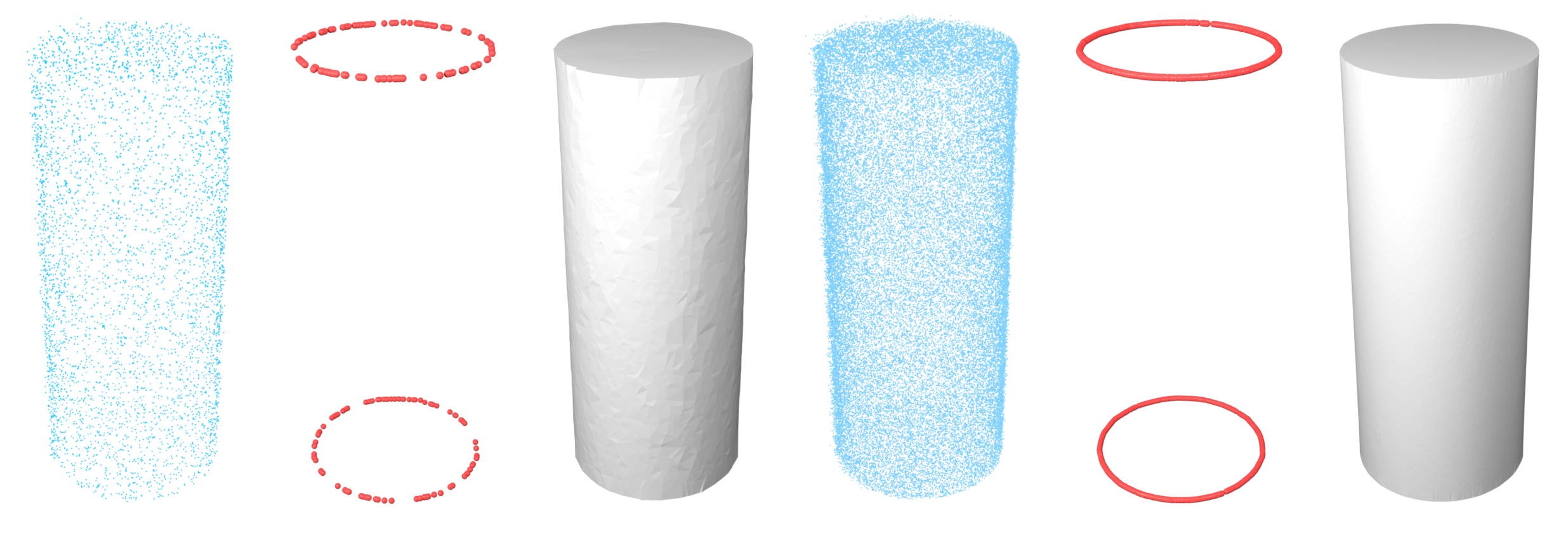}
\put(6.5,-1.5){\textbf{(a)}}
\put(23.4,-1.5){\textbf{(b)}}
\put(39.6,-1.5){\textbf{(c)}}
\put(56.7,-1.5){\textbf{(d)}}
\put(73.4,-1.5){\textbf{(e)}}
\put(90.4,-1.5){\textbf{(f)}}
\end{overpic}
\vspace{-4mm}
\caption{
\SQ{The planarity assumption in our algorithm
may not help with increasing the surface smoothness, unlike those point upsampling algorithms. 
(a-c) A 5K-size point cloud, the augmented point set and the reconstruction result. (d-f) 50K-size.}}
\label{FIG:FailureCase_yuanzhu}
\vspace{-4mm}
\end{figure}
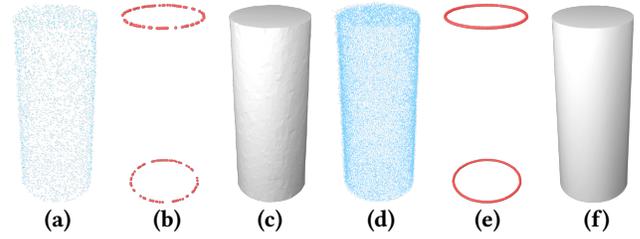

\paragraph{Planarity assumption}
\SQ{
The main purpose of the planarity assumption 
is to generate edge points and help restore feature lines. 
We use the example shown in Figure~\ref{FIG:FailureCase_yuanzhu}
to test if our algorithm can restore a smooth surface for a cylindrical shape.
Although our algorithm contains a step of denoising point locations,
the resulting points are not placed in an orderly arrangement along the generatrix; See Figure~\ref{FIG:FailureCase_yuanzhu}(a-c).
The surface smoothness has to depend on high point sampling density;
see Figure~\ref{FIG:FailureCase_yuanzhu}(d-f).
Therefore, it can be seen from Figure~\ref{FIG:FailureCase_yuanzhu} that
the additional points produced by our algorithm 
may not help with increasing the smoothness, unlike those point upsampling algorithms. 
}

\section{Conclusion}
In this paper, we propose to transform noisy point data of a CAD model into a feature-line equipped polygonal surface. 
Our algorithm consists of multiple stages,
two of which are edge-point consolidation 
and feature-line preserving reconstruction. 
For the stage of edge-point consolidation,
we propose a formulation of discrete optimal mass transport
to identify the edge zone and generate sufficiently many additional points that align with line-type geometric features. 
For the stage of feature-line preserving reconstruction,
we use  the restricted power diagram to 
interpolate the augmented point set while giving higher priority to the connections between edge points. 
Experimental results show that the combination of the two proposed techniques
is able to  exploit  the prior knowledge about CAD models, that is,
the target surface consists of multiple smooth patches stitched together by rigid feature lines. 
Tests
on both synthetic and raw-scan data
validate the effectiveness and usefulness of the proposed algorithm. 

\begin{acks}
The authors would like to thank the anonymous reviewers for their valuable comments and suggestions. This work is supported by National Key R\&D Program of China (2021YFB1715900), National Natural Science Foundation of China (62272277, 62072284) and NSF of Shandong Province (ZR2020MF153).
\end{acks}